\documentclass{article}
\usepackage{graphicx} 
\usepackage{xspace}
\usepackage[a4paper,left=2.73cm,right=2.7cm,top=3cm,bottom=3.5cm]{geometry}
\usepackage{amsfonts,amsmath,amssymb,tabstackengine}
\usepackage[normalem]{ulem}
\usepackage{tikz}
\usepackage{array}
\allowdisplaybreaks
\usepackage{slashed}
\usepackage{tensor}
\usepackage{upgreek}
\usepackage{comment}

\usepackage{cite}

\definecolor{Gray}{gray}{0.95}

\usepackage[colorlinks=true,linktocpage=true,linkcolor=blue,citecolor=blue]{hyperref}
\usepackage{graphicx}

\addtocontents{toc}{\protect\setcounter{tocdepth}{2}}
\numberwithin{equation}{section}
\usepackage{upgreek}
\usepackage{mathrsfs}
\usepackage{dsfont}
\global\long\def\ri{\mathrm{i}}%
\global\long\def\dd{\mathrm{d}}

\newcommand{\revA}{\protect\rotatebox[origin=c]{180}{A}\xspace}

 \usepackage [latin1]{inputenc}

 \usepackage{url}

\begin{document}

\begin{titlepage}
	\thispagestyle{empty}
	\begin{center}
	    { \LARGE{\bf On the Stability of Warped Anti-de Sitter Spaces\\[1ex] in Supergravity}}
        
		\vspace{50pt}
		
			{S.~Maurelli$^{1}$, R.~Noris$^{2}$, M.~Oyarzo$^{3}$, A. Saha$^4$, H. Samtleben$^5$ and M.~Trigiante$^{1}$}
		
		\vspace{25pt}

		{
		$^1${\it  Department of Applied Science and Technology, Politecnico di Torino, Corso Duca degli Abruzzi, 24, 10129 Torino, Italy\\
        INFN, Sezione di Torino, Via P. Giuria 1, 10125 Torino, Italy}

		\vspace{5pt}

		$^2${\it CEICO, Institute of Physics of the Czech Academy of Sciences,
        Na Slovance 2, 182 21 Prague 8, Czech Republic.}

		\vspace{5pt}

        $^3${\it Departamento de F\'isica de Part\'iculas \& Instituto Galego de F\'isica de Altas Enerx\'ias (IGFAE), Universidade de Santiago de Compostela, E-15782 Santiago de Compostela, Spain}
		}
        
        	\vspace{5pt}
            
		$^4${\it George P. \& Cynthia Woods Mitchell  Institute
        for Fundamental Physics and Astronomy,\\
        Texas A\&M University, College Station, TX 77843, USA}
        
		\vspace{5pt}
        
		$^5${\it ENS de Lyon, CNRS, LPENSL, UMR5672, 69342, Lyon cedex 07, France;\\
        Institut Universitaire de France (IUF)}
		
		\vspace{40pt}
		
		{ABSTRACT}
	\end{center}
    
We discuss the perturbative stability of warped anti-de Sitter (WAdS$_3$) spaces in three-dimensional gauged supergravity. To this end, we first derive the analogue of the Breitenlohner-Freedman bound  for a single scalar field on such a curved background. We then apply this analysis to the half-maximal $D=3$ gauged supergravity obtained from the compactification of the $D=6$, $\mathcal{N}=(1,1)$ supergravity on $\rm{S}^3$. Families of three-dimensional configurations, uplifting to spaces of the form $\mathrm{WAdS}_3 \times \mathrm{WS}^3$, are found by turning on appropriate vector and scalar fields, supporting the AdS$_3$ and S$^3$ warpings, respectively. Finally, we show that the spectra of scalar fluctuations satisfy the stability criteria for each solution.

\end{titlepage}
\date{January 2025}

\tableofcontents

\section{Introduction}

The issue of stability has long played a central role in the analysis of classical solutions of General Relativity. Analysing the stability of an exact solution typically requires a detailed study of the spectrum of linearized fluctuations under suitable boundary conditions. This problem has been investigated in a broad variety of settings, ranging from asymptotically flat black-hole spacetimes to anti-de Sitter vacua in gauged supergravity, where both perturbative and non-perturbative stability criteria have been developed \cite{Regge:1957td,Zerilli:1970se,Moncrief:1974am,Kodama:2003jz}.

A remarkable simplification occurs whenever the background preserves some supersymmetry within a supergravity embedding. Supersymmetric solutions, namely spacetimes admitting globally defined Killing spinors, obey BPS bounds and positive-energy theorems, implying that they extremize the energy within a given sector \cite{Schon:1979rg,Witten:1981mf,Nester:1981bjx,Townsend:1984iu,Skenderis:1999mm,Freedman:2003ax}. Under appropriate boundary conditions, this provides a strong indication of perturbative stability, 
and may furthermore exclude non-perturbative instabilities.
This makes supersymmetric vacua particularly robust starting points for holographic constructions and for the study of deformations away from maximally symmetric backgrounds.

In the holographic context, the stability of anti-de Sitter backgrounds is of central interest. The classic work of Breitenlohner and Freedman established that scalar fields in AdS may possess negative mass squared while remaining stable, provided their masses satisfy the celebrated Breitenlohner--Freedman (BF) bound \cite{Breitenlohner:1982bm,Breitenlohner:1982jf,Mezincescu:1984ev}. The analysis is performed by considering linear perturbations of a scalar field propagating on a fixed AdS background, reducing the problem to quantum field theory in curved spacetime. Stability is  determined by analysing the spectrum of normalisable solutions of the Klein--Gordon equation subject to appropriate boundary conditions, and identifying whether exponentially growing modes are present. In the dual conformal field theory, the BF bound translates into the requirement that the scaling dimensions of the corresponding operators are real and satisfy the appropriate unitarity bounds, so that bulk stability is reflected in the unitarity of the boundary theory.

A natural question is whether analogous stability criteria exist for geometries that are closely related to AdS while possessing a smaller isometry group. Among the most prominent examples are three-dimensional warped anti-de Sitter (WAdS$_3$) spacetimes. These can be viewed as one-parameter deformations of AdS$_3$ preserving only one of the two $\mathrm{SL}(2,\mathbb{R})$ factors of the original isometry group, breaking the other one down to U(1). 
They are characterized by a Ricci tensor of the form
\begin{equation}
     R_{\mu\nu}=\mathfrak{a} \,g_{\mu\nu}+\mathfrak{b} \, A_\mu  A_\nu\,,
\label{eq:Ricci-intro}
\end{equation}
with a Killing vector $A_\mu$ of constant norm, and two constants $\mathfrak{a}, \mathfrak{b}$ specifying the geometry.
These spaces have attracted considerable attention in three-dimensional gravity, where they arise as exact solutions of Topologically Massive Gravity (TMG), of three-dimensional supergravity, as well as in higher-dimensional supergravity and string theory \cite{Anninos:2008fx,Bobev:2011qx,El-Showk:2011euy,Deger:2013yla,Andrianopoli:2023dfm,Eloy:2023acy,Andrianopoli:2024twc,Eloy:2024lwn,Deger:2024xnd,Bieliavsky:2024vry,Deger:2024obg}. Perturbative stability of these backgrounds has been partially investigated in TMG \cite{Anninos:2009zi}. 
The holographic duals of these backgrounds correspond to deformations of CFTs, so-called dipole or warped CFTs \cite{Anninos:2008fx,El-Showk:2011euy,Guica:2010sw,Song:2011sr,Detournay:2012pc,Afshar:2015wjm}. 
More recently, new families of supersymmetric warped AdS$_3$ backgrounds have been obtained as near-horizon limits of Type IIB brane configurations, together with continuous non-supersymmetric deformations interpolating away from the BPS locus \cite{Maurelli:2025iba,Maurelli:2025ueo}.

The main goal of this paper is to provide a systematic approach to describe  warped AdS$_3$ geometries within three-dimensional supergravity and to analyze their stability within this framework. Unlike AdS$_3$, warped AdS geometries possess a reduced symmetry group and a modified asymptotic structure, features that may substantially alter the spectrum of fluctuations and the admissible boundary conditions. 
As a first step, we follow the strategy successfully employed in the AdS case and analyze linear perturbations of a minimally coupled scalar field propagating on fixed warped AdS$_3$ backgrounds. We consider all three Lorentzian warped geometries---lightlike, spacelike and timelike --- treating the metric as non-dynamical and neglecting the backreaction of the scalar fluctuation. This allows us to isolate the simplest sector of the perturbation problem, determine the spectrum of scalar modes, analyze the role of boundary conditions, and derive general criteria for perturbative stability.

As a first step, we follow the strategy successfully employed in the AdS case and analyze a minimally coupled massive probe scalar field propagating on each warped AdS$_3$ background. This allows us to determine the spectrum of scalar modes for a given mass, analyze the role of boundary conditions, and derive general criteria for perturbative stability that will later be applied when we perturb all the fields. Next, we give a systematic account of the description of WAdS$_3$ solutions in general three-dimensional gauged supergravity. In the AdS$_3$ case, different AdS solutions in a given matter-coupled gravitational theory correspond to different expectation values of the scalar fields, which are determined as the extremal points of the scalar potential. The AdS length is then read off from the value of the potential at the extremal point. 
In the case of warped AdS solutions, due to the reduced isometry group, also the vector fields of the theory can acquire non-vanishing values on the background, proportional to the Killing vector $A_\mu$ from (\ref{eq:Ricci-intro}). We show that the full set of equations of motion for the warped case reduces to the algebraic problem of finding the extremal points of a larger {\em generalized scalar potential} which captures the scalar and the vector degrees of freedom.
We then linearize the general equations of motion around a given background, as a starting point for the stability analysis of fluctuations. 

Eventually, we apply the perturbation analysis to concrete examples: WAdS$_3$ solutions in a specific half maximal $D=3$ gauged supergravity, which uplift to the $D=6$, ${\cal N}=(1,1)$ model. Solutions to the latter, of the form $\mathrm{WAdS}_3 \times \mathrm{WS}^3$, WS$^3$ being the warped sphere, were constructed in \cite{Maurelli:2025iba}. In the spacelike and timelike case, they are non-supersymmetric, for generic values of the warpings of AdS$_3$ and S$^3$ and would naturally correspond to the $D=3$ warped backgrounds studied here.
The linearized equations are recast in terms of a system of coupled Schr\"odinger equations for stationary states. The parameters of these equations are expressed in terms of the embedding tensor, defining the gauging in three-dimensions, as well as the two Fourier modes along the boundary directions of the warped background. For a special class of lightlike $\mathrm{WAdS}_3$ backgrounds, corresponding to Schr\"odinger spaces with dynamical coefficient $n=2$, as well as for squashed spacelike and stretched timelike $\mathrm{WAdS}_3$ geometries, we prove that perturbative stability is achieved, provided the three-dimensional scalar masses satisfy a modified BF bound. The latter is defined by an effective mass value which depends on the momentum along the boundary direction of the $\mathrm{WAdS}_3$ spacetime, through the warping parameter. In particular, the warped BF bound is lower compared to the known one. We conclude that, for such spaces, the warping improves perturbative stability, at least as far as the effective $D=3$ supergravity truncation is concerned. 

The rest of this paper is organized as follows: in section \ref{sec2} we review the main mathematical properties of WAdS$_3$ spacetimes and provide an explicit coordinate description of their metrics and contact structures, for each warping. In section \ref{sec3}, we discuss stability of massive scalar Klein-Gordon modes on a general $\mathrm{WAdS}_3$ background. We reduce the problem to the analysis of solutions of a corresponding characteristic Schr\"odinger problem for stationary states. In section
\ref{sec4}, we present the general setup in which $\mathrm{WAdS}_3$ solutions are described within a three-dimensional gauged supergravity and spell out the linearized field equations around these solutions. In section \ref{sec5}, we focus on a $\mathcal{N} = 8, D=3$ model coupled to four vector multiplets, and provide explicit solutions for each warping. In section \ref{sec6}, we derive the corresponding scalar fluctuation masses, comparing the results with the criteria developed in section \ref{sec3}. Finally, in section \ref{sec7}, we summarize our results and discuss possible directions for future work.

\section{The Geometry of ${\rm WAdS}_3$ Spacetimes}\label{sec2}
In this section, we review the main properties of WAdS$_3$ spaces, while simultaneously defining our conventions. For an in-depth review of these geometries we refer the reader to \cite{Bieliavsky:2024vry}.

Let us recall that the metric $g_{\mu\nu}$ of a WAdS$_3$ spacetime can be obtained by deforming the anti-de Sitter one $g^{(0)}_{\mu\nu}$ as follows:
\begin{equation}
    g_{\mu\nu}=g^{(0)}_{\mu\nu}+\varpi\,\bar{A}_\mu \bar{A}_\nu\,,
\end{equation}
where $\bar{A}_\mu \equiv g^{(0)}_{\mu\nu}\,K^\nu$, $K^\nu$ being one of the ${\rm SO}(2,2)$ ${\rm AdS}_3$ Killing vectors, and $\varpi$ is a real parameter. The characteristic feature of these spaces is that the Ricci tensor has the form
\begin{equation}
       R_{\mu\nu}=\mathfrak{a}(\varpi)\,g_{\mu\nu}+\mathfrak{b}(\varpi) \bar A_\mu \bar A_\nu\,.
\end{equation}
By setting the warping parameter $\varpi$ to zero, one recovers the anti-de Sitter geometry, which corresponds to the values $\mathfrak{b}=0,\,\mathfrak{a}=-2/L^2$, $|L|$ being the anti-de Sitter radius. Depending on whether $\bar{A}_\mu$ is spacelike, lightlike or timelike the warped spacetime will be referred to as spacelike, lightlike or timelike warped AdS$_3$ and be denoted by ${\rm WAdS}_3^+,\,{\rm WAdS}_3^0,\,{\rm WAdS}_3^-$, respectively. These spaces are known as $\eta$-Einstein, K-contact, the latter property being related to the fact that the 1-form $\bar{A}=\bar{A}_\mu dx^\mu$, with the exception of the lightlike case \cite{Murcia:2019cck}, is a Killing contact form on the manifold.
This in turn follows from the fact that $\bar{A}$ satisfies a Beltrami-like equation of the form
\begin{equation}
    \star \dd \bar A=\mathfrak{m}\,\bar A\,,
\end{equation}
where $\star$ is the Hodge duality operation on the three-dimensional spacetime. For the anti-de Sitter space, the \emph{Beltrami mass} is $\mathfrak{m}=\pm 2/L$. In general, the sign of $\mathfrak{m}$ defines a chirality that depends on whether the Killing vector $K^\mu$ defining $\bar{A}_\mu$, belongs to the $\mathfrak{sl}(2,\mathbb{R})_L$ or $\mathfrak{sl}(2,\mathbb{R})_R$ part of the anti-de Sitter isometry algebra.
The three parameters $\mathfrak{a}$, $\mathfrak{b}$, $\mathfrak{m}$ are related, in all cases, by the following equation:
\begin{equation}
\mathfrak{m}^2+2\,\mathfrak{a}+2\,\mathfrak{b}\,|\bar{A}|^2 = 0.
\end{equation}
Below we list the explicit description of these spaces, employed in the rest of the paper, and make a specific choice for the aforementioned chirality. 

\paragraph{Metric and $\eta$-Einstein structure of spacelike warped $\mathrm{AdS}_3$ (WAdS$_3^+$)}
\begin{align}\label{WAdS3+ metric}
    \dd s^2&=\frac{L^2}{4z^2}\left(-\dd t^2+\dd z^2\right)+\lambda\left(\dd x-\frac{L\dd t}{2z}\right)^2\,,\qquad x^\mu=(t,z,x)\,,\nonumber\\
    \bar A_\mu \dd x^\mu &= \kappa \left(\dd x-\frac{L\dd t}{2z}\right)\,, \qquad 
    \bar A^\mu \partial_\mu = \frac{\kappa}{\lambda}\partial_x\,,\qquad |A|^2=\frac{\kappa^2}{\lambda}\,,\nonumber\\ 
    \mathfrak{a}&=\frac{2(\lambda-2)}{L^2}\,, \qquad \mathfrak{b}=\frac{4\lambda(1-\lambda)}{L^2 \kappa^2}\,,\qquad \mathfrak{m}=\frac{2\sqrt{\lambda}}{L}\,.
\end{align}

\paragraph{Metric and $\eta$-Einstein structure of lightlike warped $\mathrm{AdS}_3$ (WAdS$_3^0$)}
\begin{align}\label{parametersWAdS0}
    \dd s^2&=\frac{L^2}{z^2}\left(\dd x_+\dd x_-+\dd z^2\right)-\alpha L^4\frac{\dd x_-^2}{z^4}\,,\qquad x^\mu=(x_+,z,x_-)\,,\nonumber\\
    \bar A_\mu \dd x^\mu &= \kappa \frac{\dd x_-}{z^2}\,, \qquad 
    \bar A^\mu \partial_\mu = \frac{2\kappa}{L^2}\partial_{x_+}\,,\qquad |A|^2=0\,,\nonumber\\
    \mathfrak{a}&=-\frac{2}{L^2}\,, \qquad \mathfrak{b}=\frac{4 L^2 \alpha}{ \kappa^2}\,,\qquad \mathfrak{m}=\frac{2}{L}\,.
\end{align}

\paragraph{Metric and $\eta$-Einstein structure of timelike warped $\mathrm{AdS}_3$ (WAdS$_3^-$)}
\begin{align}\label{parametersWAdS-}
    \dd s^2&=\frac{L^2}{4z^2}\left(\dd x^2+\dd z^2\right)-\lambda\left(\dd t-\frac{L\dd x}{2z}\right)^2\,,\qquad x^\mu=(t,z,x)\,,\nonumber\\
    \bar A_\mu \dd x^\mu &= \kappa \left(\dd t-\frac{L\dd x}{2z}\right)\,, \qquad 
    \bar A^\mu \partial_\mu = -\frac{\kappa}{\lambda}\partial_t\,,\qquad |A|^2=-\frac{\kappa^2}{\lambda}\,,\nonumber\\ 
    \mathfrak{a}&=\frac{2(\lambda-2)}{L^2}\,, \qquad \mathfrak{b}=\frac{4\lambda(\lambda-1)}{L^2 \kappa^2}\,,\qquad \mathfrak{m}=\frac{2\sqrt{\lambda}}{L}\,.
\end{align}

\section{Perturbative stability of WAdS$_3$}\label{sec3}
In this section, we investigate the stability of a real massive Klein--Gordon scalar field, propagating on a $\mathrm{WAdS}_{3}$ background. The analogous problem for $\mathrm{AdS}_{d+1}$ spacetimes was originally studied in \cite{Breitenlohner:1982bm,Breitenlohner:1982jf} and later revisited in \cite{Townsend:1984iu,Mezincescu:1984ev}, leading to the well-known Breitenlohner--Freedman (BF) stability bound
\begin{equation}\label{BFbound}
    m^{2}\geq m^2_{\mathrm{BF}}= - \frac{d^2}{4 L^{2}}\,,
\end{equation}
where $m$ represents the mass of the scalar field and $L$ is the AdS radius. A convenient strategy to establish this bound, which we shall adopt here, consists in reducing the Klein--Gordon equation to a one-dimensional Schr\"odinger problem for stationary modes, supplemented by appropriate boundary conditions. In this context, a way of establishing perturbative stability of a background is requiring that every admissible solution of the Schr\"odinger problem satisfies the following two requirements:
\begin{itemize}
    \item[1)] \emph{Perturbative consistency}: the scalar perturbation must remain bounded with respect to the background geometry and therefore must not exhibit exponential growth in spacetime \cite{Moroz:2009kv}.
    \item[2)] \emph{Energy stability}: the corresponding conserved energy must be bounded from below.
\end{itemize}
Such an approach is well suited for backgrounds which admit a foliation in constant-time slices, allowing to define a conserved energy. These include standard AdS$_3$ and its lightlike and spacelike deformations. The timelike warping deserves a separate discussion; we will address this in section \ref{stabilitytimelike}. 

We begin by reviewing the standard stability analysis for $\mathrm{AdS}_{3}$, which provides the basis for deriving the corresponding BF bounds in both lightlike and spacelike warped AdS$_3$ backgrounds and then address the perturbative stability of timelike solutions.

\subsection{Stability of AdS$_{3}$}\label{sec31}
We begin here by considering the AdS$_3$ metric in the Poincar\'e patch,
\begin{align}\label{metricAdS3}
    \dd s^2=L^{2}\bigg(\frac{-\dd t^2+ \dd x^2+\dd z^2}{z^2}\bigg) \,,
\end{align}
written in terms of coordinates $x^\mu=(t,x,z)$. The hypersurface $z=0$ corresponds to the AdS$_3$ boundary, while the limit $z\rightarrow\infty$ describes the interior of such spacetime. A real massive scalar fluctuation $\phi$ on this background is governed by the following action and equation of motion 
\begin{align}
    S&=-\int \dd^3x \sqrt{g}\left(\nabla_\mu \phi\nabla^\mu \phi+m^2 \phi^2  \right)\,,\label{KGaction}\\
    0&=\Box\phi-m^2\phi\,,\label{KGeom}
\end{align}
where $m$ is the mass of the fluctuation. Given the action \eqref{KGaction}, one can construct the following conserved current
\begin{align}
    \mathcal J_\mu&=T_{\mu\nu}\xi^\nu= T_{\mu t}\,,\label{Isocurrent}
\end{align}
where 
\begin{align}
    T_{\mu\nu}&=\partial_{(\mu}\phi\partial_{\nu)}\phi-\frac12g_{\mu\nu}(g^{\rho\sigma}\partial_\rho\phi\partial_\sigma\phi+m^2\phi^2)\,
\end{align}
is the stress--energy tensor of the scalar field and $\xi=\partial_t$ is the timelike Killing vector. We choose the boundary of the integration region to be $\partial M=\Sigma_{t_1}\cup \bar\Sigma_{t_2}\cup \Sigma_{z_1}\cup \bar \Sigma_{z_2}$, $\Sigma_t$ being a spacelike slice at time $t$ and $\Sigma_z$ being a timelike slice at fixed $z$. The conservation law then implies
\begin{align}\label{divergence}
    0=\int_M \dd ^3 x\sqrt{g}\,\nabla_\mu \mathcal J^\mu =\left(\int_{\Sigma_{t_2}} -\int_{\Sigma_{t_1}}  \right)\dd^2 x\sqrt{g}\,n_\mu \mathcal J^\mu+\left(\int_{\Sigma_{z_2}}-\int_{\Sigma_{z_1}}\right)\dd ^2 x \sqrt{g}\,n_\mu \mathcal J^\mu\,,
\end{align}
where $n_\mu=\partial_\mu t$ on $\Sigma_t$ and $n_\mu=\partial_\mu z$ on $\Sigma_z$. After integrating by parts the time and radial derivatives in the corresponding boundary contributions, the surface terms from the constant time and radial slices cancel identically, so that the above condition can be cast in the following form 
\begin{align}\label{conservation0}
    0&=-E(t_2) +E(t_1)+ \Delta(z_2)-\Delta(z_1)\,,
\end{align}
where 
\begin{align}\label{energyandflux}
    E(t) &= -\int_{\Sigma_t}\dd x^2\sqrt{g}\,n_\mu \mathcal J^\mu - \int_{\partial \Sigma_t}\!\!\dd x\,\frac{L}{2z}\phi\partial_z\phi\,,\qquad 
    \Delta(z) = \int_{\Sigma_z}\dd x^2\sqrt{g}\,n_\mu \mathcal J^\mu -\int_{\partial \Sigma_z}\!\!\dd x\,\frac{L}{2z}\phi\partial_z\phi\,.
\end{align}
In the following, we shall take $z_1=0$ and $z_2=\infty$, covering the full Poincar\'e patch. Hence, the quantity $E(t)$, to be interpreted as the total energy of the system, is conserved if and only if
\begin{align}\label{conservation}
    \Delta(z=\infty)-\Delta(z=0)=0\,,
\end{align}
namely if the net energy flux across the radial boundary vanishes. 

Since the spacetime is invariant under translations along $t$ and $x$, it is natural to decompose the scalar field into Fourier modes, that is
\begin{align}\label{expansion}
    \phi(t,x,z)=\int \frac{\dd \omega \dd p}{(2\pi)^2}e^{-\ri \omega t+\ri p x}\sqrt{\frac zL}\,\psi_{\omega,p}(z)\,,
\end{align}
where $\omega,p\in \mathbb R$. The Klein-Gordon equation \eqref{KGeom} reduces to a 1-dimensional Schr\"odinger equation for the stationary ``wave function'' $\psi_{\omega,p}(z)$ 
\begin{align}\label{SchAdS3}
    -\psi''_{\omega,p}(z)+V(z)\psi_{\omega,p}(z)=\mathcal E_{\mathrm{Sch}}\psi_{\omega,p}(z)\,,
\end{align}
with potential and Schr\"odinger energy, respectively
\begin{align}\label{potentialand energy}
    V(z)=\frac{L^2m^2+\frac34}{z^2}\,,\qquad \mathcal E_{\mathrm{Sch}}=\omega^2-p^2\,
\end{align}
and satisfying $\psi_{\omega,p}=\psi^\dagger_{-\omega,-p}$, by the reality of $\phi$. Equation \eqref{SchAdS3} reduces a Bessel equation, whose general solution reads 
\begin{align}\label{schsolads}
    \psi_{\omega,p}(z)=\sqrt{\frac zL}\Big(A_{\omega,p} J_\nu(\sqrt{\mathcal E_{\mathrm{Sch}}} z)+B_{\omega,p} Y_\nu (\sqrt{\mathcal E_{\mathrm{Sch}}} z)\Big)\,,
\end{align}
where 
\begin{align}
    \nu^2=L^2m^2+1=L^2(m^2-m^2_{\mathrm{BF}})\,.
\end{align}
Here $J_{\nu}(z)$ and $Y_{\nu}(z)$ are the Bessel functions of first and second kind, respectively. Using the Fourier mode decomposition, the conserved energy can be expressed entirely in terms of the radial wavefunctions,
\begin{align}\label{Energy0}
    E(t)&=\frac14\int_{z_1}^{z_2} \dd z\int \frac{\dd \omega\dd \omega' \dd p}{(2\pi)^3}e^{-\ri (\omega-\omega')t}(\omega+\omega')^2\psi_{\omega,p} \psi^\dagger_{\omega',p}\,,
\end{align}
which is conserved provided \eqref{conservation} holds, where 
\begin{align}
    \Delta(z)&=-\frac\ri4\int_{t_1}^{t_2}\dd t\int \frac{\dd \omega\dd \omega' \dd p}{(2\pi)^3}e^{-\ri (\omega-\omega')t}(\omega+\omega')(\psi_{\omega,p} \partial_z\psi^\dagger_{\omega',p}-\psi^\dagger_{\omega',p} \partial_z \psi_{\omega,p})\,.
\end{align}
Since energy conservation must hold for arbitrary values of $t_1$ and $t_2$, the above boundary condition is satisfied if 
\begin{align}
    \int \frac{\dd \omega\dd \omega' \dd p}{(2\pi)^3}e^{-\ri (\omega-\omega')t}(\omega+\omega')\left.(\psi_{\omega,p} \partial_z\psi^\dagger_{\omega',p}-\psi^\dagger_{\omega',p} \partial_z \psi_{\omega,p})\right\vert_{z=0}^{z=\infty}=0\,,
\end{align}
which clearly holds if the following stronger condition is satisfied
\begin{align}\label{boundarycond}
    \left.(\psi_{\omega,p} \partial_z\psi^\dagger_{\omega',p}-\psi^\dagger_{\omega',p} \partial_z \psi_{\omega,p})\right\vert_{z=0}^{z=\infty}=0\,.
\end{align}
Remarkably, the above equation is equivalent to the orthogonality of the Schr\"odinger eigenfunctions \cite{PhysRev.80.797}:
\begin{align}
    0&=\left.(\psi_{\omega,p} \partial_z\psi^\dagger_{\omega',p}-\psi^\dagger_{\omega',p} \partial_z \psi_{\omega,p})\right\vert_{z=0}^{z=\infty}=(\omega^2-\omega'^2)\int_{0}^{\infty} \dd z\, \psi_{\omega,p}(z) \psi^\dagger _{\omega',p}(z)\,.\label{orthogonality}
\end{align}
For the class of solutions under consideration, perturbative consistency is automatically guaranteed, since the spacetime dependence is purely oscillatory. On the other hand, as mentioned above, energy stability holds for those solutions which have a finite, positive and conserved energy. The latter condition is achieved by configurations which vanish at the boundary $z=0$, obtained by setting $B_{\omega,p}=0$. The orthogonality relation is then satisfied by virtue of \eqref{leon} in appendix \ref{app0}, where the following expression of the energy of the fluctuation is derived
\begin{align}\label{Energy1}
    E(t)&=\int \frac{\dd \omega \dd p}{(2\pi)^2}|\omega| |A_{\omega,p}|^2=\mathrm{const.}>0\,.
\end{align}

In order to assess the perturbative stability of the AdS$_3$ background, we need to consider all solutions to the Schr\"odinger problem and check if both stability conditions are satisfied. For this reason, one needs to establish if there are solutions with $\omega=-\ri \beta$ and $\nu=\ri \gamma$ that satisfy the boundary conditions and ensure conservation of the corresponding energy. They clearly exhibit exponential growth in time\footnote{The corresponding reality condition now reads $\psi_{\beta,p}=\psi_{\beta,-p}^\dagger$.} and satisfy a Schr\"odinger equation with potential as in \eqref{potentialand energy} and $\mathcal E_{\mathrm{Sch}}=-\beta^2-p^2$. Leaving the details to appendix \ref{app0}, the solution which satisfies the said requirements has the form 
\begin{align}\label{Hankel1}
    \psi_{\beta,p}=A_{\beta,p}\sqrt{\frac zL}H^{(1)}_{\ri \gamma}(\ri \sqrt{|\mathcal E_{\mathrm{Sch}}|} z)\,,
\end{align}
where $H_{\ri \gamma}^{(1)}(z)$ is the Hankel function of first kind, and $\beta$ and $p$ satisfy the following quantization condition for the Schr\"odinger energy  
\begin{align}\label{quantisationE}
    |\mathcal E_{\mathrm{Sch}}|\equiv \mathcal E_n= e^{\frac{2\pi n}{\gamma}}\mathcal E_0\,, \qquad n\in \mathbb Z\,,\qquad \mathcal E_0>0\,.
\end{align}
The corresponding conserved energy reads
\begin{align}\label{Energy3}
    E(t)=-\frac14 \int_0^\infty \dd z\int\frac{\dd \beta\dd \beta' \dd p}{(2\pi)^3} e^{-(\beta+\beta')t}(\beta-\beta')^2\psi_{\beta,p}\psi^\dagger_{\beta',p}\,
\end{align}
and it can be shown to be negative, if we restrict to states for which $\psi_{\beta,p}=\psi_{-\beta,p}$. Moreover, by restricting to an infinite selection of states in the spectrum, \eqref{Energy3} can be proven to be unbounded from below, as shown in appendix \ref{app0}. We have therefore identified admissible solutions to the Schr\"odinger problem, which violate the perturbative consistency requirement as well as the energy criterion. For this reason, scalar fields with $\nu^2<0$, that is $m^2<-\frac{1}{L^2}$, destabilize the AdS$_3$ background. One can verify that solutions with positive $\nu^2$ and imaginary $\omega$ do not satisfy the required orthogonality relations. Indeed, in this regime, \eqref{hankelmagic} would feature a $\sinh$ in place of a $\sin$, which only vanishes when its argument is zero, that is when $\mathcal{E}_{\mathrm{Sch}}=\mathcal{E}_{\mathrm{Sch}}'$.

This concludes the review of the derivation of the BF bound for the AdS$_3$ case. In the following, we will apply the same strategy to derive the corresponding results for spacelike and lightlike WAdS$_3$ backgrounds.

\subsection{Stability of WAdS$_3^+$}
Let us consider here the spacelike warped AdS$_3$ case, whose metric and $\eta$-Einstein structure can be found in \eqref{WAdS3+ metric}. We will refer to a \emph{squashed} deformation when the deforming parameter $\lambda$ belongs to the interval $0< \lambda \leq 1$, while \emph{stretched} deformation will be used to denote the complementary range $\lambda> 1$. The undeformed case corresponds to the value $\lambda =1$. 

In the following, we will focus on the squashed deformation, since it features a timelike vector $\partial_{t}$, defined on the entire patch. We will then interpret it as describing a consistent time evolution in order to define the energy. This allows us to follow the same derivation, based on Stokes' theorem, as in equations \eqref{divergence} and \eqref{conservation0}. Following the results obtained for the AdS$_3$ case, conservation of $E(t)$ requires the vanishing of the radial energy flux \eqref{conservation}.

The translational symmetries of the background allow us to decompose the scalar field into Fourier modes. To analyze the stability of WAdS$_3^+$, we shall directly focus on Fourier modes of the Klein-Gordon scalar field with imaginary frequency
\begin{align}
    \phi(t,x,z)=\int \frac{\dd \beta \dd p}{(2\pi)^2}e^{-\beta t+\ri p x}\,\psi_{\beta,p}(z)\,.
\end{align}
The Klein-Gordon equation then reduces to a corresponding 1-dimensional problem with the following potential and Schr\"odinger energy 
\begin{align}
    V(z)=\frac{L^2}{4z^2}\left(m^2+p^2\left(\frac{1}{\lambda}-1\right)\right)-\frac{\ri \beta L p}{z}\,, \qquad \mathcal E_{\mathrm{Sch}}=-\beta^2\,.
\end{align}
The effect of the spacelike warping deformation is encoded in the effective Schr\"odinger potential through an additional positive contribution proportional to $p^2$. The associated solution is written in terms of Whittaker functions \cite{Gradshteyn:1943cpj} \footnote{Useful properties of the Whittaker functions can also be found in \cite{NIST:Whittaker}.}
\begin{align}
    \psi_{\beta,p}(z)=A_{\beta,p}W_{\frac{\ri}{2} L p \,\mathrm{Sign}\beta,\nu}(2|\beta| z)+B_{\beta,p}M_{\frac{\ri}{2} L p \,\mathrm{Sign}\beta,\nu}(2|\beta| z)\,,
\end{align} 
where we defined 
\begin{align}\label{nuspacelike}
    \nu=\nu(p)=\frac12\sqrt{L^2m^2+1+L^2p^2\left(\frac{1}{\lambda}-1\right)}\,.
\end{align}
For squashed deformations, the reality of $\nu$ is guaranteed by $m^2\geq m^2_{\mathrm{BF}}$. The parameter becomes purely imaginary if instead $m^2<m^2_{\mathrm{BF}}$ and $p\in [-p_0,\,p_0]$, where $p_0=\sqrt{\frac{\lambda  \left(L^2 m^2+1\right)}{(\lambda -1) L^2}}$.

The corresponding energy
\begin{align}\label{energyspacelike}
    E(t)=-\frac{\sqrt\lambda}{4}\int \dd z \int \frac{\dd\beta\dd\beta'\dd p}{(2\pi)^3}e^{-(\beta+\beta')t}\left((\beta'-\beta)^2+\frac{\ri L p}{z}(\beta'-\beta)\right)\psi^\dagger_{\beta',p}\psi_{\beta,p}\,
\end{align}
is conserved provided 
\begin{align}\label{conservationenergy}
    0=\int \frac{\dd\beta\dd\beta'\dd p}{(2\pi)^3}e^{-(\beta+\beta')t}(\beta'-\beta)\left.\left(\psi_{\beta,p}\partial_z\psi^\dagger_{\beta',p}-\psi^\dagger_{\beta',p}\partial_z\psi_{\beta,p}\right)\right\vert^\infty_0\,.
\end{align}
We seek for an infinite set of states satisfying the above condition. For this reason, we restrict to imaginary values of $\nu$, i.e. $\nu=\ri \gamma$, and impose $B_{\beta,p}=0$, in such a way that the solution vanishes in the interior $z\to \infty$. The above boundary condition further requires $\beta$ to satisfy the following quantization relation
\begin{align}\label{betaspectrum}
    |\beta|\equiv \beta_n=e^{\frac{2\pi n}{\gamma}}\beta_0\,, \qquad n\in \mathbb Z\,, \qquad \beta_0>0\,,
\end{align}
together with $A^\dagger _{\beta',p}A_{\beta,p}=A^\dagger _{\beta',-p}A_{\beta,-p}$, for values $\beta,\beta'$ within the spectrum and with same sign, as shown in the appendix \ref{app0}, equation \eqref{orthowhittaker}. 

The Schr\"odinger equation implies the following equality 
\begin{align}
    \left((\beta'-\beta)+\frac{\ri L p}{z} \right)\psi^\dagger_{\beta',p}\psi_{\beta,p}=\frac{1}{\beta+\beta'}\partial_z(\psi_{\beta,p}\partial_z\psi^\dagger_{\beta',p}-\psi^\dagger_{\beta',p}\partial_z\psi_{\beta,p})\,,
\end{align}
which can be used to express the energy \eqref{energyspacelike} in terms of the orthogonality relation 
\begin{align}\label{energyspacelike2}
    E(t)=-\frac{\sqrt\lambda}{4}\int \frac{\dd\beta\dd\beta'\dd p}{(2\pi)^3}e^{-(\beta+\beta')t}\frac{\beta'-\beta}{\beta'+\beta}\left.\left(\psi_{\beta,p}\partial_z\psi^\dagger_{\beta',p}-\psi^\dagger_{\beta',p}\partial_z\psi_{\beta,p}\right)\right\vert^\infty_0\,.
\end{align}
One can verify that the only non-vanishing contributions to the energy come from $\beta'=-\beta$ and that, as shown in the appendix \ref{app0}, there exists a suitable range of $p$, around $p=0$, such that the energy is negative and unbounded from below. We have therefore found an infinite set of states which satisfy the chosen boundary conditions and violate both criteria for perturbative stability. A crucial ingredient in the construction of these states is the condition $\nu^2<0$. For real $\nu$, the orthogonality condition cannot be satisfied, precisely as in the AdS case. If the scalar mass is assumed to be independent of the momentum $p$, requiring stability for all Fourier modes reduces the condition to the original AdS$_3$ BF bound, $m^2\geq -\frac{1}{L^2}$. However, as we shall see in later sections, this assumption is not generally valid for the WAdS$_3$ solutions arising in supergravity. In such backgrounds, the linearized scalar fluctuations can acquire an effective momentum-dependent mass, allowing for configurations satisfying
$-\frac{1}{L^2}-\frac{(1-\lambda) p^2}{\lambda }\leq m^2(p)\leq-\frac{1}{L^2}$. For this reason, we can reformulate the stability condition in terms of $\nu^2\geq0$, rather than directly in terms of the scalar mass.

\subsection{Stability of WAdS$_3^0$}
The metric for the lightlike warped AdS$_3$ geometry, given in \eqref{parametersWAdS0}, can be obtained as a deformation of the AdS$_3$ metric in the Poincar\'e patch.
In the following, we shall assume the on/off parameter $\alpha$ to be positive, $\alpha>0$, implying that the vector $\partial_-$ is always timelike, which in turn ensures that $x_-$ defines a consistent time evolution. Although the hypersurfaces of constant $x_-$ are not standard spacelike Cauchy surfaces, since their normal vectors are null, Stokes' theorem can still be applied to derive a conserved charge and the associated boundary condition. We follow the same derivation based on Stokes' theorem, applied to the conservation of the new current $\mathcal J^\mu=T^\mu{}_-$, as in equations \eqref{divergence} and \eqref{conservation}. We define $E(x_-)$ and $\Delta(z)$ as in \eqref{energyandflux}, where $n_\mu=\partial_\mu x_-$ on $\Sigma_{x_-}$. As in the AdS$_3$ case, conservation of $E(x_-)$ requires the vanishing of the radial energy flux \eqref{conservation}.
Exploiting the translational symmetries of the background, we use the Fourier decomposition introduced in \eqref{expansion}, where $\omega$ and $p$ are now conjugate to $x_-$ and $x_+$, respectively. The corresponding conserved energy takes the form
\begin{align}\label{energyWAdS0}
    E(x_-)=\frac12\int_{0}^{\infty}\dd z \int\frac{\dd \omega\dd\omega'\dd p}{(2\pi)^3}e^{-\ri (\omega-\omega')x_-}p(\omega+\omega')\psi_{\omega,p}\psi^\dagger_{\omega',p}\,,
\end{align}
whose conservation requires, as before, 
\eqref{boundarycond}. The Klein-Gordon equation in turn reduces to a Schr\"odinger one \eqref{SchAdS3} with 
\begin{align}
    V(z)=\frac{L^2m^2+\frac34+4\alpha p^2L^2}{z^2}\,,\qquad \mathcal E_{\mathrm{Sch}}=4\omega p\,.
\end{align}
The deforming parameter $\alpha$ affects the Schr\"odinger potential by a contribution proportional to $p^2$. Moreover, differently from the AdS$_3$ case, the Schr\"odinger energy is no longer quadratic in the Fourier momenta but depends linearly on $\omega$ and $p$, as observed in \cite{Moroz:2009kv}. The most general solution is written as in \eqref{schsolads}, where now 
\begin{align}\label{nulightlike}
    \nu=\nu(p)=\sqrt{L^2m^2+1+4\alpha p^2L^2}\,.
\end{align}
We can distinguish two qualitatively different classes of solutions to the Schr\"odinger problem, as in the undeformed case. The first one corresponds to the regime where both Schr\"odinger energy and $\nu^2$ are positive. In this case, one can consider $B_{\omega,p}=0$, and restrict to modes with positive Schr\"odinger energy. In order to satisfy the orthogonality condition, we can use again \eqref{leon}, while positivity and finiteness of \eqref{energyWAdS0} are achieved by suitably choosing the wave packets. The second set of solutions is related to a complementary regime, where both the Schr\"odinger energy and $\nu^{2}$ are negative. The latter condition requires $m^2L^2\le-1=m_{{\rm BF}}^2\,L^2$ and is achieved by restricting to $p\in [-p_0,\,p_0]$, where
$p_0^2=-(m^2 L^2+1)/(4\alpha L^2)$. These solutions are defined by
\begin{align}
    B_{\omega,p}=\ri  A_{\omega,p}\,,\qquad A_{\omega,p}=\theta(-4\omega p)a_{\omega,p}\,, \qquad \nu=\ri \gamma\,.
\end{align}
Explicitly, the full configuration reads\footnote{Notice that the condition $\omega p<0$ imposed by the Heaviside step function allows for both positive and negative values of $\omega$ and $p$. This ensures that one can always impose the reality condition on $\phi$.}
\begin{align}\label{solutionWAdS0}
    \phi(x_{-},x_{+},z)=\int \frac{\dd \omega \dd p}{(2\pi)^2}e^{-\ri \omega x_{-}+\ri p x_{+}}\theta(-4\omega p)a_{\omega,p}\frac zL H^{(1)}_{\ri \gamma}(\ri \sqrt{|\mathcal E_{\mathrm{Sch}}|} z)\,,
\end{align}
where the integral in $p$ is restricted to the interval defined above.
Despite the apparent similarity with \eqref{alternativesol}, there is an important difference: the above solution remains bounded in time and it therefore provides a sensible perturbation \cite{Moroz:2009kv}. We can therefore check if the boundary conditions are satisfied and compute the corresponding energy. Adapting  \eqref{hankelmagic} to the new definitions of $\gamma$ and $\mathcal E_{\mathrm{Sch}}$, we obtain a different, yet analogous, quantization condition for the Schr\"odinger energy
\begin{align}
    \mathcal E_{\mathrm{Sch}}=e^{\frac{2\pi n}{\gamma}}\mathcal E_0\,, \qquad n\in \mathbb Z\,,\qquad \mathcal E_0<0\, \quad \implies\quad  \omega_n = e^{\frac{2\pi n}{\gamma}}\omega_0\,.
\end{align}
Being $\mathcal E_{\mathrm{Sch}}<0$, we  therefore restrict to a set of states for which $\omega_0>0, p<0$ and $\omega_0 <0, p>0$. The resulting energy is negative and can be made arbitrarily large in magnitude by an appropriate choice of the wave packet coefficients $a_{\omega,p}$, as shown in appendix \ref{app0}, equation \eqref{Energy0final}. These solutions provide an explicit example where the perturbative analysis is consistent, i.e. perturbations do not diverge in time, but the corresponding energy is unbounded from below. According to the criteria previously described, these states, which correspond to $\nu^2<0$, should be excluded via a suitably modified BF bound. The inclusion of the energy criterion therefore changes the stability assessment relative to \cite{Moroz:2009kv}, where all scalar perturbations of WAdS$_3^0$ were found to be stable.
Similarly to the AdS$_3$ case, the orthogonality conditions are not satisfied for the remaining values of Schr\"odinger energy and $\nu$. 

In the following, we shall formulate stability of WAdS$_3^0$ in terms of the positivity of $\nu^2$, for the same reasons outlined in the previous subsection, implying
\begin{align}
    \nu^2\geq 0\qquad \iff\qquad m^2(p)\geq -\frac{1}{L^2}-4\alpha p^2\,.
\end{align}

\subsection{Stability of WAdS$_3^-$}\label{stabilitytimelike}
Spacetimes of this kind deserve a somewhat separate treatment in our analysis. In their squashed version $\lambda<1$, they are free of closed timelike curves (CTC) and a global foliation in constant-time slices is possible. In spite of this regularity feature, however, it was shown in \cite{Israel:2004vv} that squashed WAdS$_3^-$ cannot be embedded in superstring theory as pure NS-NS backgrounds. In fact, to our knowledge, no known solution to  superstring theory or $D=10$ supergravity contains a squashed WAdS$_3^-$ factor\footnote{Notice that our WAdS$_3^-$ metric corresponds to the first line of eq. (2.8) of \cite{Hoare:2022asa}, in which our deformation parameter $\lambda$ coincides with $1+\kappa^2$. The supersymmetric solutions with timelike fibration, given in that reference, feature a real $\kappa$ and thus correspond, in our terminology, to \emph{stretched} warpings, in contrast to the terminology used in the same reference.}. Stretched WAdS$_3^-$, on the other hand, are 3D sections of a generalization of the G\"odel spacetime \cite{Reboucas:1982hn}, and share with the latter the presence of CTCs. This makes the definition of a conserved energy problematic. We shall refrain from trying to define a consistent conserved energy on these spaces\footnote{See \cite{Donnay:2015joa} for a definition of conserved charges on these spaces within NMG. } and thus from applying to them the approach followed in the previous sections, to address the issue of perturbative stability. We shall rather use a different approach and only require the condition of perturbative consistency. This, as we shall see, will suffice to rule out squashed timelike metrics and, for the stretched ones, provide a condition analogous to that found for the other warped geometries.

Let us start by considering a scalar fluctuation, satisfying the Klein--Gordon equation, on this class of backgrounds, using for the latter the global coordinate patch introduced in \cite{Balasubramanian:2003kq,Anninos:2009jt}. In this case, the metric takes the form 
\begin{align}\label{timelikeglobal}
\dd s^{2} & =\frac{L^{2}}{4}\left[(1+r^{2})\dd u^{2}+\frac{\dd r^{2}}{1+r^{2}}-\lambda(\dd\tau-r\dd u)^{2}\right]\,.
\end{align}
As mentioned above, it is straightforward to verify that a timelike foliation exists for $\lambda<1$, with time direction $\tau$. By decomposing a generic scalar field into Fourier modes as
\begin{align}\label{timelikesol}
    \phi(\tau,u,r)=\int \frac{\dd \tilde\omega\,\dd \tilde p}{(2\pi)^2}e^{-\ri \tilde \omega \tau +\ri\tilde p u}\tilde{\psi}_{\tilde\omega,\tilde p}(r)\,,
\end{align}
the Klein--Gordon equation reduces to a radial one
\begin{align} \label{eq-timelike-global}
    \tilde{\psi}''_{\tilde\omega,\tilde p}+\frac{2 r}{r^2+1}\tilde{\psi}'_{\tilde\omega,\tilde p}-\left(\frac{\lambda  L^2 m^2-4 \tilde{\omega} ^2}{4 \lambda (1+ r^2)}+\frac{(\tilde p-r \tilde\omega )^2}{\left(r^2+1\right)^2}\right)\tilde{\psi}_{\tilde\omega,\tilde p}=0\,.
\end{align}
Introducing the parameter 
\begin{align}
    \tilde{\nu}  &= \frac12\sqrt{L^2m^2+1+4\tilde\omega^2\left(1-\frac{1}{\lambda}\right)} \, ,
\end{align}
the requirement that the solution be regular at both $r\to+\infty$ and $r\to-\infty$ leads to the quantization condition \cite{Anninos:2009jt}
\begin{align}\label{quantcond}
    -\tilde{\omega}   \pm \left( n + \frac{1}{2} + \tilde{\nu} \right) = 0\,, \qquad n\in\mathbb N\,.
\end{align}
In particular, finiteness of the scalar fluctuation requires $\tilde\nu$ to be real. If $0<\lambda<1$, for sufficiently large values of $\tilde\omega$, the parameter $\tilde\nu$ becomes imaginary and the perturbation grows exponentially in time. This leaves only the regime $\lambda>1$ as the only compatible with perturbative stability. As mentioned above, however, this regime features CTC and thus it is not quite physical. 

It is now useful to relate the reality of $\tilde \nu$ to the description of WAdS$_3^-$ in \eqref{parametersWAdS-}. In those coordinates, the scalar field admits the Fourier decomposition
\begin{align}
    &\phi(t,z,x)=\int \frac{\dd \omega\,\dd  p}{(2\pi)^2}e^{-\ri  \omega t +\ri p x}{\psi}_{\omega, p}(z)\,,
\end{align}
which leads to the following radial Schr\"odinger equation
\begin{align}
    &\psi''_{\omega, p}-\frac{L^2}{4z^2}\left(m^2+\omega^2\left(1-\frac{1}{\lambda}\right)\right)\psi_{\omega, p}+\frac{L p \omega }{z}\psi_{\omega, p}-p^2\psi_{\omega, p}=0\,.
\end{align}
The corresponding parameter is
\begin{align}\label{nutimelike}
    &{\nu}  = \frac12\sqrt{L^2m^2+1+L^2\omega^2\left(1-\frac{1}{\lambda}\right)} \,.
\end{align}
Since in the $r\to \infty$ limit \eqref{timelikeglobal} reduces to \eqref{parametersWAdS-}, up to a rescaling of the time coordinate $$\tau=\frac 2L t\,,\qquad \tilde\omega=\frac{L}{2}\omega\,,$$ the solutions to the corresponding radial equations of motion must coincide. This allows us to identify the two parameters, in this regime, as
\begin{align}
    \tilde \nu = \nu \,.
\end{align}
Due to this identification, the parameter $\nu$ inherits the reality requirement on $\tilde \nu$: we will therefore use such condition as a criterion for the perturbative stability of WAdS$_3^-$.

\paragraph{A statement about stretched WAdS$_3^+$:}
In the above analysis, we have given evidence for the instability of squashed WAdS$_3^-$. As shown in \cite{Hoare:2022asa}, a spacelike warped AdS background can be obtained through a double Wick rotation of a timelike one, combined with a suitable analytic continuation of the warping parameter. In particular, this procedure maps squashed WAdS$_3^-$ into stretched WAdS$_3^+$. Therefore our analysis of the former points towards the instability of the latter, as well.

\section{Gauged Supergravity in $D=3$}\label{sec4}
Having provided criteria to establish perturbative stability for all deformations of AdS$_3$, we now turn to the study of $\rm{WAdS_{3}}$ solutions within the setting of a general  $D=3$ gauged supergravity model \cite{deWit:2003ja}. The bosonic sector of the latter consists of, besides the metric, $n$ scalar fields $\phi^i$, spanning a scalar manifold $\mathcal{M}_{\mathrm{scal}}$ with isometry group $G$ and $n_v$ vector fields $A_\mu^{\mathcal M}$, transforming in the adjoint representation of $G$. Its combined action on scalars and vectors, makes $G$ a global symmetry of the ungauged limit of the theory. The model features a gauge group $\mathcal G\subset G$, defined by an embedding tensor $\Theta_{\mathcal M\mathcal N}=\Theta_{\mathcal N\mathcal M}$ (for reviews see \cite{Samtleben:2008pe,Trigiante:2016mnt}) , defining the gauge generators $X_\mathcal M$, as linear combinations of the isometry generators $t_\mathcal M$
\begin{align}
    X_{\mathcal M}=\Theta_{\mathcal M\mathcal N}t^\mathcal N\,.
\end{align}
These generators satisfy a linear condition (depending on the number of supercharges), as well as a universal quadratic one 
\begin{align}
    [X_\mathcal M,X_\mathcal N]=-X_{\mathcal M\mathcal N}{}^\mathcal P X_{\mathcal P}\,,
\end{align}
where the structure constants of the gauge algebra are defined in terms of the adjoint representation $X_{\mathcal M\mathcal N}{}^\mathcal P$ of the gauge generators.
The bosonic lagrangian has the following general form
\begin{equation}
    \mathcal{L} = \sqrt{-g}\left[ \mathcal{R} - \frac{1}{2}\mathcal{D}_{\mu}\phi^{i}\mathcal{D}_{\nu}\phi^{j}\mathcal{G}_{ij}(\phi)g^{\mu\nu} - V(\phi)\right] + \mathcal{L}_{CS}\,,
\end{equation}
where $\mathcal{G}_{ij}$ is the metric on the scalar manifold  $\mathcal{M}_{\mathrm{scal}}$, which defines the non-linear $\sigma$-model action for the scalar fields and the scalar potential $V(\phi)$ is completely defined by the embedding tensor \cite{deWit:2003ja}. The Chern-Simons lagrangian reads
\begin{equation}
    \mathcal{L}_{CS} = -\varepsilon^{\mu\nu\rho}\Theta_{\mathcal{M\mathcal{N}}}A_{\mu}^{\mathcal{M}}\left( \partial_{\nu}A_{\rho}^{\mathcal{N}} + \frac{1}{3}X_{\mathcal{P}\mathcal{Q}}{}^{\mathcal{N}}A_{\rho}^{\mathcal{P}}A_{\nu}^{\mathcal{Q}}\right)\,,
\end{equation}
where the Levi-Civita symbol is defined as $\varepsilon_{012}=-\varepsilon^{0 1 2}=1$. Finally, the gauge-covariant derivatives of the scalar fields read
\begin{equation}
    \mathcal{D}_{\mu}\phi^{i} = \partial_{\mu}\phi^{i} + A_{\mu}^{\mathcal{M}}\Theta_{\mathcal{M}\mathcal{N}}\tilde{K}^{\mathcal{N}i} = \partial_{\mu}\phi^{i} +A_{\mu}^{\mathcal{M}}K_{\mathcal{M}}^{i} \,,
\end{equation}
where $\tilde{K}^{\mathcal{N}i}$ are the Killing vectors on $\mathcal M_{\mathrm{scal}}$. 

The corresponding bosonic field equations are then computed to be 
\begin{equation}\label{EE3D}
    R_{\mu\nu}-g_{\mu\nu}V(\phi)- \frac{1}{2}\mathcal{D}_{\mu}\phi^{i}\mathcal{D}_{\nu}\phi^{j}\mathcal{G}_{ij}(\phi) =0\,,
\end{equation}
\begin{equation}\label{Veq3D}
    \sqrt{-g}K_{\mathcal{M}}^{i}\mathcal{D}^{\mu}\phi^{j}\mathcal{G}_{ij}(\phi) + \varepsilon^{\mu\nu\rho}\Theta_{\mathcal{M}\mathcal{N}}F_{\nu\rho}^{\mathcal{N}} =0\,,
\end{equation}
\begin{equation}\label{eom3D}
   \nabla_{\mu}( \mathcal{G}_{ij} \mathcal{D}^{\mu}\phi^{j}) - \frac{1}{2}\partial_{i}\mathcal{G}_{jk}(\phi)\mathcal{D}_{\mu}\phi^{j}\mathcal{D}^{\mu}\phi^{k}-\partial_{i}K_{\mathcal{M}}^{j}A_{\mu}^{\mathcal{M}}\mathcal{G}_{jk}\mathcal{D}^{\mu}\phi^{k} - \partial_{i}V(\phi) =0\,,
\end{equation}
where $\nabla_{\mu}$ is Levi-Civita covariant derivative of the three-dimensional spacetime and the non-abelian field strengths of the vector fields have the following general form
\begin{align}
    F^{\mathcal{M}}_{\mu\nu} &= \partial_{\mu}A_{\nu}^{\mathcal{M} } - \partial_{\nu}A_{\mu}^{\mathcal{M} } + X_{\mathcal{P}\mathcal{Q}}{}^{\mathcal{M}}A_{\mu}^{\mathcal{P}}A_{\nu}^{\mathcal{Q}}\,.
\end{align}

We now look for solutions of the model describing $\rm{W}AdS_{3}$ spaces. Since these metrics are obtained by deforming $\mathrm{AdS}_3$ by means of a 1-form $\bar{A}=\bar A_\mu \dd x^\mu$, dual to one of the $\mathrm{AdS}_3$ Killing vectors, a necessary minimal ingredient for defining such background is that one of the vectors $\bar A^\mathcal M_\mu$ be non-zero on the solution and provide the 1-form $\bar A$, defining the warping. Throughout the paper, barred quantities will be evaluated on the considered background. We shall therefore use the following $G$-covariant ansatz for the vector fields on the solution
\begin{align}
    \bar A^\mathcal M_\mu= C^\mathcal M \bar A_\mu\,, \label{Abar}
\end{align}
where $C^\mathcal M$ is a constant vector, formally transforming in the adjoint representation of $G$.
Finally, we require the scalar fields to be constant on such solution
\begin{align}
    \phi^i\equiv \bar\phi^i\,.
\end{align} 
The equations of motion, evaluated on the ansatz defined above, reduce to 
\begin{align}
     0&=(\mathfrak{a} - \bar V) \bar g_{\mu\nu} + (\mathfrak{b} -\mathcal{\bar K})\bar A_{\mu}\bar A_{\nu}\,, \label{eq01}\\
     0&=C^{\mathcal{M}}(\mathcal{\bar K}_{\mathcal{M}\mathcal{N}} + 2 \mathfrak{m}\,\Theta_{\mathcal{M}\mathcal{N}})\bar A^{\mu}\,, \label{eq02} \\
    0&=\left.\partial_{i}\left(|A|^2 \mathcal{ K} +  V(\phi)\right)\right\vert_{\phi=\bar\phi}\,, \label{eq03}
\end{align}
where we defined $\mathcal{K}\equiv \frac{1}{2}C^{\mathcal{M}}C^{\mathcal{N}}\mathcal{K}_{\mathcal{M}\mathcal{N}}$ and $\mathcal{K}_{\mathcal{MN}}\equiv K_{\mathcal{M}}^{i}\mathcal{G}_{ij}K_{\mathcal{N}}^{j}$\,.  For non-vanishing values of $|\bar A|^2$, that is for spacelike and timelike warped $\mathrm{AdS}_3$ solutions, the second and third equations of motion can be rewritten in a more compact form as the following ``attractor equations''
\begin{align}
    0&=\left.\frac{\partial\mathscr{V}}{\partial\Phi^{X}}\right\vert_{\Phi=\bar\Phi}\,,\label{generalized extr}
\end{align}
in terms of a \emph{generalized scalar potential}
\begin{equation}
    \mathscr{V} = V(\phi) + \frac{1}{2}|A|^2 C^{\mathcal{M}}C^{\mathcal{N}}\left(\mathcal{K}_{\mathcal{M}\mathcal{N}} +2 \mathfrak{m} \Theta_{\mathcal{M}\mathcal{N}}\right)
\end{equation}
and a set of generalized fields $\Phi^{X}=\{\phi^{i},C^{\mathcal{M}}\}$.
The above equations should then be solved for the constants values of $\mathfrak{a}$, $\mathfrak{b}$, $\mathfrak{m}$, $\bar\phi^i$ and $C^\mathcal M$.

\subsection{Linearized Field Equations}
Given a solution of the equations \eqref{EE3D}, \eqref{Veq3D} and \eqref{eom3D}, one can expand the dynamical fields around such background as
\begin{align}
        g_{\mu\nu} &= \bar{g}_{\mu\nu} + \epsilon\, h_
        {\mu\nu} + \mathcal{O}(\epsilon^2)\,,\nonumber\\
        A_{\mu}^{\mathcal{M}} &= \bar{A}_{\mu}^{\mathcal{M}} + \epsilon\,\delta A_{\mu}^{\mathcal{M}} + \mathcal{O}(\epsilon^2)\,,\nonumber\\
        \phi^{i} &=\bar{\phi}^{i} + \epsilon\,\delta\phi^{i} + \mathcal{O}(\epsilon^2)\,.
\end{align}
The linearized equations of motion then read
\begin{align}
    0&=R_{\mu\nu}^{(1)} - \bar V h_{\mu\nu} -\bar{\mathcal{K}}_{\mathcal{MN}}\bar{A}_{(\mu}^{\mathcal{M}}\delta A_{\nu)}^{\mathcal{N}} -\bar{K}_{\mathcal{M}}^{i} \bar{\mathcal{G}}_{ij}\bar{A}_{(\mu}^{\mathcal{M}}\partial_{\nu)}\delta\phi^{j} - \left(\overline{\partial_{i} V}\bar{g}_{\mu\nu} + \overline{\partial_{i}\mathcal{K}}\bar{A}_{\mu}\bar{A}_{\nu}\right)\delta\phi^{i}\,, \label{EEq pert}\\
    0&=\left(\frac12 h \bar{g}_{\mu\nu} - h_{\mu\nu}\right)\bar{A}^{\mathcal{N}\nu}\bar{\mathcal{K}}_{\mathcal M\mathcal N} +\bar{\mathcal{K}}_{\mathcal M\mathcal N} \bar{g}_{\mu\nu}\delta A^{\mathcal{N}\nu} + \bar{K}_{\mathcal{M}}^{i}\mathcal{\bar G}_{ij}\partial_{\mu}(\delta\phi^{j}) + \bar A_\mu^{\mathcal{N}}\overline{\partial_{k}\mathcal{K}_{\mathcal{MN}}}\delta\phi^{k} \nonumber \\
    &+ 2\sqrt{\bar{g}}\,\varepsilon_{\mu\nu\rho}\Theta_{\mathcal{MN}}\big[ \partial^{\nu} \delta{A^{\mathcal{N}\rho}} + X_{\mathcal{PQ}}{}^{\mathcal{N}}\bar{A}^{\mathcal{P}\nu}\delta{A^{\mathcal{Q}\rho}}\big]\,, \label{Veq pert}\\
    0&=\bar{\mathcal {G}}_{ij}\bar{\nabla}^\mu\left(\bar{K}^j_{\mathcal M}(-h_{\mu\nu}\bar{A}^{\mathcal M\nu}+\frac12h \bar{A}^{\mathcal M}_\mu+\delta A_\mu^{\mathcal M})+\partial_\mu\delta \phi^j\right)- \partial_{i,j}\left.\Big(|\bar A|^2\mathcal { K}+V\Big)\right\vert_{\phi=\bar\phi}\delta\phi^{j}\nonumber\\
    &+2\overline{ D_iK^m_{\mathcal M}}\mathcal{\bar G}_{mn}\bar A^{\mathcal M\mu}\left(K^n_{\mathcal N}\left(\frac12h_{\mu\nu}\bar A^{\mathcal N \nu}-\delta A^{\mathcal N}_\mu\right)-\partial_\mu\delta\phi^n\right)\,,\label{scal pert}
\end{align}
where $D_{i}$ is the covariant derivative on the scalar manifold, and the first-order perturbation of the Ricci tensor is given by 
\begin{align}
    R_{\mu\nu}^{(1)} = \frac{1}{2}\left(\bar{\nabla}_{\alpha}\bar{\nabla}_{\mu}h^{\alpha}{}_{\nu} +\bar{\nabla}_{\alpha}\bar{\nabla}_{\nu}h^{\alpha}{}_{\mu}  - \bar{\nabla}_{\alpha}\bar{\nabla^{\alpha}}h_{\mu\nu} -\bar{\nabla}_{\mu}\bar{\nabla}_{\nu}h \right)\,.
\end{align}
In order to assess the perturbative stability of the background, we shall focus on the independent scalar fluctuations, obtained after fixing the residual gauge freedom. These will consist in the projection of the vector fluctuations along $\bar A_\mu$, as well as the remaining non-Goldstone scalar fields. These independent fluctuations will be shown to satisfy a coupled system of Schr\"odinger equations, which, once diagonalized, will be compared with the analysis performed in section \ref{sec3}. In what follows we shall apply this analysis to a specific model. 

\section{A Half-Maximal $D=3$ Gauged Model}\label{sec5}
In \cite{Maurelli:2025iba,Maurelli:2025ueo,Maurelli:2026dpp}, supersymmetric and non-supersymmetric Type IIB solutions of the form ${\rm WAdS}_3\times {\rm WS}^3\times T^4$ were found within the half-maximal $D=6$, $\mathcal{N} = (1,1)$ ungauged supergravity. We wish to study the perturbative stability of such backgrounds by looking for $\rm{WAdS_{3}}$ solutions within a suitable $D=3$ gauged model, originating from the compactification of the former six-dimensional theory on a three-sphere. In what follows, we will then focus on a particular gauging of this three-dimensional, half-maximal supergravity, denoted by  \revA  in \cite{Eloy:2021fhc}.

The scalar manifold of the higher-dimensional model is
\begin{equation}
    \mathcal{M}_{\mathrm{scal}}^{(D=6)}  = {\rm SO}(1,1)\times\frac{\mathrm{SO}(4,n)}{\mathrm{SO}(4)\times \mathrm{SO}(n)}\,,
\end{equation} 
where $n$ is the number of six-dimensional vector multiplets, making the total number of vector fields $4+n$. The field content of the model is completed by a gravitational multiplet which, besides other fields, also includes an antisymmetric rank-2 tensor field. The scalar manifold of the three-dimensional supergravity model is 
\begin{equation}
    \mathcal{M}_{\mathrm{scal}} = \frac{\mathrm{SO}(8,4+n)}{\mathrm{SO}(8)\times \mathrm{SO}(4+n)}\,,
\end{equation}
and the gauge group has the following form:
$$\mathcal{G}_{{\rm gauge}} = \left({\rm SO}(4)\ltimes T^6\right)\times T^{4+n}\,,$$
where SO$(4)$ is the isometry group of S$^3$ and $T^6,\,T^{4+n}$ are translational isometries. As opposed to  \cite{Eloy:2021fhc}, where the $n=0$ case was studied, we will choose here $n=4$. 

A useful approach is to consider the following subgroup of the isometry group of the scalar manifold:
\begin{equation}
    {\rm SO}(3,3)\times {\rm SO}(1,1)\times {\rm SO}(4,4)\subset {\rm SO}(8,8)\,,\label{subgroup}
\end{equation}
with respect to which the isometry generators decompose as follows:
\begin{align}
    {\rm Adj}_{{\rm SO}(8,8)}\,\rightarrow &\,\left({\rm Adj}_{{\rm SO}(3,3)},{\bf 1}\right)_0\oplus  \left({\bf 1},{\bf 1}\right)_0\oplus ,\left({\bf 1},\,{\rm Adj}_{{\rm SO}(4,4)}\right)_0\oplus ({\bf 6},{\bf 8})_{0}\oplus ({\bf 1},{\bf 8})_{+1}\oplus ({\bf 1},{\bf 8})_{-1}\oplus\nonumber\\
   & \oplus  ({\bf 6},{\bf 1})_{+1}\oplus  ({\bf 3},{\bf 1})_{-1}\,,
\end{align}
where the grading refers to the ${\rm O}(1,1)$ factor in \eqref{subgroup}.
The generators of the ${\rm SO}(4)$ subgroup of $\mathcal{G}_{{\rm gauge}}$ span the maximal compact subgroup of ${\rm SO}(3,3)$, while the factors $T^6$ and $T^8$ are generated by the isometries in $({\bf 6},{\bf 1})_{+1}$ and in $({\bf 1},{\bf 8})_{+1}$, respectively. The latter are central charges of the gauge algebra. The fundamental representation ${\bf 16}$ of ${\rm SO}(8,8)$, whose basis elements will be labeled by the  indices ${M,\,N}=1,\dots,\,16$, branches under $ {\rm SO}(3,3)\times {\rm SO}(1,1)\times {\rm SO}(4,4)$ as
\begin{align}\label{funddec}
    {\bf 16}&\rightarrow\,({\bf 6,1})_0 \oplus {\bf (1,1)}_{+1}\oplus  {\bf (1,1)}_{-1}\oplus {\bf (1,8)}_{0}\,,\nonumber\\
    V_{{M}}&=(V_m,\,V^m,\,V_0,\,V^0,\,V_{\mathtt{I}},\,V^\mathtt{I})\,,
\end{align}
where $m=1,2,3$ and $\mathtt{I}=1,\dots, 4$. In this basis, the invariant, symmetric tensor $\eta_{MN}$ of the fundamental representation has the following non-vanishing entries: 
\begin{align}
    \eta_{m}{}^n=\eta^n{}_m=\delta^n_m\,,\,\,\eta_{0}{}^0=\eta^0{}_{0}=1\,,\,\,\,\eta_\mathtt{I}{}^\mathtt{J}=\eta^\mathtt{J}{}_\mathtt{I}=\delta_\mathtt{I}^\mathtt{J}\,.
\end{align}
The generators $J_{\mathcal{M}}\equiv J_{MN}$ of ${\rm SO}(8,8)$ are labeled by the antisymmetric couple $[MN]$, to be denoted by $\mathcal{M}=1,\dots, 120$. In the fundamental representation, they read
\begin{equation}
    (J_{MN})_P{}^Q=2\,\eta_{P[M}\,\delta_{N]}^Q\,,
\end{equation}
and close the commutation relations
\begin{equation}
    [J_{MN},\,J_{PQ}]=-4 \eta_{[P[M}J_{N]Q]}\,.
\end{equation}
With these conventions, the generators in the $({\bf 1},{\bf 8})_{+1}$ are proportional to $J^{0\mathtt{I}},\,J^0{}_\mathtt{I}$, while those in  $({\bf 6},{\bf 1})_{+1}$ are expressed in terms of the matrices $J^{0,m},\,J^0{}_m$. \par
We adopt the solvable parametrization of the coset in which the scalar manifold is described as isometric to a solvable Lie group generated by the Borel subalgebra $\mathfrak{s}$ of $\mathfrak{so}(8,8)$. 
The scalar fields are then parameters of this solvable Lie algebra, which is generated by the 8 non-compact Cartan generators and by the shift generators corresponding to the positive roots.
The parametrization is also defined by the prescription for the exponentiation of the solvable Lie algebra element, which defines the coset representative $\mathcal{V}$.
An orthogonal basis of Cartan generators is
$$H_m=-J_m{}^m\,,\quad H_4=-J_0{}^0\,, \quad H_{4+\mathtt{I}}=-J_\mathtt{I}{}^\mathtt{I}\,.$$
It is useful to describe the nilpotent generators corresponding to the positive roots in the Borel subalgebra with respect to the subgroup
\begin{equation}
    {\rm SL}(3,\mathbb{R})\times {\rm SO}(1,1)'\times {\rm SO}(1,1)\times {\rm SO}(4,4)\subset  {\rm SO}(3,3)\times  {\rm SO}(1,1)\times {\rm SO}(4,4)\,.
\end{equation}
The Borel subalgebra $\mathfrak{s}$ reads
\begin{equation}\nonumber
    \mathfrak{s}= \mathfrak{s}[\mathfrak{sl}(3,\mathbb{R})]\oplus \mathfrak{s}[\mathfrak{so}(4,4)]\oplus \mathfrak{o}(1,1)'\oplus \mathfrak{o}(1,1)\oplus ({\bf 3,1})_{(1,1)}\oplus ({\bf 3',1})_{(0,2)}\oplus  ({\bf 1,8})_{(1,0)}\oplus  ({\bf 3,8})_{(0,1)}\oplus  ({\bf 3,1})_{(-1,1)}\,,
\end{equation}
where $\mathfrak{s}[\mathfrak{sl}(3,\mathbb{R})]$ and $\mathfrak{s}[\mathfrak{so}(4,4)]$ are the Borel subalgebras of $\mathfrak{sl}(3,\mathbb{R})$ and $\mathfrak{so}(4,4)$, respectively.
The two $\mathfrak{o}(1,1)$-generators $H,\,H'$, which the gradings in subscript refer to, are
\begin{equation}
    \mathfrak{o}(1,1)={\rm Span}(H)\,,\,\,H=H_4\,;\,\,\,\,  \mathfrak{o}(1,1)'={\rm Span}(H')\,,\,\,H'=H_1+H_2+H_3\,.
\end{equation}
The coset representative is defined as follows:
\begin{align}
    \mathcal{V}=\mathcal{V}_{{\rm SO}(4,4)}\cdot \mathcal{V}_{({\bf 3,1})_{(1,1)}}\cdot \mathcal{V}_{({\bf 3',1})_{(0,2)}}\cdot \mathcal{V}_{ ({\bf 1,8})_{(1,0)}}\cdot \mathcal{V}_{({\bf 3,8})_{(0,1)}}\cdot \mathcal{V}_{\mathfrak{sl}(3,\mathbb{R})}\cdot \mathcal{V}_{({\bf 3,1})_{(-1,1)}}\cdot \mathcal{V}_{{\rm O}(1,1)^2}\,,
\end{align}
where
\begin{align}
   \mathcal{V}_{({\bf 3,1})_{(1,1)}}&=\exp(\Sigma_m\,J^{m,0})\,,\,\,  \mathcal{V}_{({\bf 3',1})_{(0,2)}}=\exp\left(\frac{1}{2}\,\sigma^m\,J^{np}\epsilon_{mnp}\right)\,,\,\,  \mathcal{V}_{({\bf 1,8})_{(1,0)}}=\exp\left(\nu_\mathtt{I}\,J^{0,\mathtt{I}}+\nu^\mathtt{I}\,J^0{}_\mathtt{I}\right)\,,\nonumber\\
   \mathcal{V}_{({\bf 3,8})_{(0,1)}}&=\exp\left(\rho_{m,\mathtt{I}}\,J^{m,\mathtt{I}}+\rho_m{}^\mathtt{I}\,J^m{}_\mathtt{I}\right) \,,\,\,\, \mathcal{V}_{({\bf 3,1})_{(-1,1)}}=\exp\left(\upphi_m\,J^m{}_0\right)\,,\,\,\,\mathcal{V}_{{\rm O}(1,1)^2}=\exp\left(\chi\,H_\chi+\xi\,H_\xi\right)\,,\nonumber\\
   \mathcal{V}_{{\rm SO}(4,4)}&\in \exp\left(\mathfrak{s}[\mathfrak{so}(4,4)]\right)\,\sim\,\frac{{\rm SO}(4,4)}{{\rm S}[{\rm O}(4)\times {\rm O}(4)]}\,.
\end{align}
Here we defined 
$$H_\chi\equiv H_1+H_2+H_3-3 H_4\,,\qquad H_\xi\equiv H_1+H_2+H_3+ H_4\,.$$
The coset representative $\mathcal{V}_{\mathfrak{sl}(3,\mathbb{R})}$ of ${\rm SL}(3,\mathbb{R})/{\rm SO}(3)$ is chosen in the following upper-triangular form:
\begin{equation}
   \left(\mathcal{V}_{\mathfrak{sl}(3,\mathbb{R})} \right)_m{}^n=\left(\begin{matrix}1 & p_1 & p_2+p_1 p_3\cr 0 & 1 & p_3\cr 0 & 0 & 1\end{matrix}\right)\cdot \left(\begin{matrix}e^{s_1}& 0& 0\cr 0 & e^{-s_1+s_2} & 0\cr 0 & 0 & e^{-s_2}\end{matrix}\right)\,.
\end{equation}
The gauge algebra has dimension 20 and it is generated by 20 independent matrices, out of the following 26 non-vanishing gauge generators $X_{\mathcal{M}}$:
$$X_{m,n}=X^{m,n}\,,\,\,\,X_m{}^n=-X_n{}^m\,,\,\,\,X_{m,0}\,,\,\,\,X^{m}{}_0\,,\,\,\,X_{0,\mathtt{I}}\,,\,\,\,X_0{}^\mathtt{I}\,.$$
Below, we give the explicit form of the 20 independent gauge generators $X_{\mathcal{M}}=X_{[MN]}$, in the chosen basis:\\
${\rm SO}(4)$-generators
\begin{align}\label{SO4gen}
    X^3{}_0+2\Lambda X^{1,2}&=\frac{1}{2\sqrt{2}}\,\left(J_{1}{}^2-J_2{}^1\right)\,,\qquad X^2{}_0-2\Lambda X^{1,3}=\frac{1}{2\sqrt{2}}\,\left(J_{3}{}^1-J_1{}^3\right)\,,\nonumber\\
    X^1{}_0+2\Lambda X^{2,3}&=\frac{1}{2\sqrt{2}}\,\left(J_{2}{}^3-J_3{}^2\right)\,,\qquad X_{1,0} -2\Lambda X_{3}{}^2=\frac{1}{2\sqrt{2}}\,\left(J_{2,3}+J^{2,3}\right)\,,\nonumber\\ 
    X_{2,0}+2\Lambda X_{3}{}^1&=-\frac{1}{2\sqrt{2}}\,\left(J_{1,3}+J^{1,3}\right)\,,\qquad X_{3,0}-2\Lambda X_2{}^1=\frac{1}{2\sqrt{2}}\,\left(J_{1,2}+J^{1,2}\right)\,,\nonumber
    \end{align}
$T^6$-generators
\begin{align}
    X^{1,2}&=\frac{1}{2\sqrt{2}}\,J^{3,0}\,,\qquad X^{1,3}=-\frac{1}{2\sqrt{2}}\,J^{2,0}\,, \qquad X^{2,3}=\frac{1}{2\sqrt{2}}\,J^{1,0}\,,\nonumber\\
    X_{2}{}^1&=-\frac{1}{2\sqrt{2}}\,J_3{}^0\,, \qquad X_{3}{}^1=\frac{1}{2\sqrt{2}}\,J_2{}^{0}\,, \qquad X_{3}{}^2=-\frac{1}{2\sqrt{2}}\,J_1{}^0\,,\nonumber
\end{align}
$T^8$-generators
\begin{align}
    X_{0,\mathtt{I}}&=\frac{\Lambda}{\sqrt{2}}\,J_\mathtt{I}{}^0\,,\qquad X_{0}{}^\mathtt{I}=\frac{\Lambda}{\sqrt{2}}\,J^{\mathtt{I},0}\,,
\end{align}
where $\Lambda$ is a parameter of the embedding tensor, which will be related to the WAdS$_3$ radius.

This model features a maximally supersymmetric ${\rm AdS}_3$ vacuum, corresponding to the $\mathrm{AdS}_3\times \mathrm{S}^3$ supersymmetric background of the $D=6$, $\mathcal{N}=(1,1)$ model and to the ${\rm AdS}_3\times \mathrm{S}^3\times T^4$ $1/2$-BPS solution of Type IIB superstring theory, which describes the near-horizon geometry of a system of D1-D5 branes (or of the dual F1-NS5 system). The ${\rm SO}(4,4)/{\rm S}[{\rm O}(4)\times {\rm O}(4)]$ submanifold of $\mathcal{M}_{\rm scal}$ is spanned by the moduli of this background, which will not play a role in our analysis. The scalar fields parametrizing the central charge generators in the $({\bf 1,8})_{+1}$ originate from the reduction of the vector fields in $D=6$ to three dimensions, upon dualization of the resulting three-dimensional 1-forms into scalars.

Vacuum solutions of this model can be uplifted to six dimensions using general formulae derived within Exceptional Field Theory (ExFT), which can be found in \cite{Eloy:2021fhc}. In this reference, classes of (non)-supersymmetric solutions of the form\footnote{The manifold $M_{\xi,\zeta}^3$ is a compact three-dimensional Riemannian manifold with the topology of $\rm{S}^{3}$, whose metric depends on two parameters. For specific loci in the space of deformations, the metric reduces to that of warped or round $\rm{S}^3$. We will discuss these cases in more detail in the next section.} ${\rm AdS}_3\times M_{\xi,\zeta}^3\times T^4$ were constructed by uplifting vacua of the $D=3$ half-maximal model, and the KK spectrum was derived. As emphasized earlier, the main objective of the present investigation is not the construction of new higher-dimensional solutions with geometry ${\rm WAdS}_3\times {\rm WS}^3\times T^4$, broad classes of which have already been engineered, but rather studying their perturbative stability.

\subsection{Background Configuration}
The purpose of this subsection is to present explicit WAdS$_3$ backgrounds of the model under consideration, by solving the equations \eqref{eq01}, \eqref{eq02}, \eqref{eq03}. In the next section, we will discuss the corresponding perturbative stability.

To start with, we need to identify the vector, among the gauge fields, which acquires a non-vanishing profile in the ${\rm WAdS}_3$ background. Following \eqref{Abar}, the latter is singled out by the constants $C^{\mathcal{M}}$, whose only non-vanishing components will be chosen to be
$C^{0,\mathtt{I}=4},C^{0}{}_{\mathtt{I}=4}$, so that the corresponding vectors gauge central charge generators in ${\bf (1,8)}_{+1}$. This choice is instrumental to our analysis as it implies the following simplification: 
\begin{equation}
    \bar{A}^{\mathcal{M}}\,X_{\mathcal{MN}}{}^{\mathcal{P}}\,X_{\mathcal{P}}={\bar A}_\mu\,C^{\mathcal{M}}\,X_{\mathcal{MN}}{}^{\mathcal{P}}\,X_{\mathcal{P}}=0\,,
\end{equation}
so that, on the background, the non-abelian terms in the vector field strengths $F^{\mathcal{M}}$, as well as in $\mathcal{D}_\mu\delta A^{\mathcal{M}}_\nu$, when contracted with $\Theta_{\mathcal{MN}}$, vanish. 

Let us collect here the variables appearing in equations \eqref{eq01}, \eqref{eq02}, \eqref{eq03}:
\begin{align}
    \{C^\mathcal M,\phi^i\}=\{C^{0,\mathtt{I}=4},C^{0}{}_{\mathtt{I}=4},\chi,\xi,\upphi_m,p_m,s_1,s_2,\rho_{m,\mathtt{I}},\rho_m{}^\mathtt{I},\nu_\mathtt{I},\nu^\mathtt{I},\sigma^m,\Sigma_m,\boldsymbol{\tau}\}\,,
\end{align}
where $\boldsymbol{\tau}$ are the moduli fields parametrizing the ${\rm SO}(4,4)/{\rm S}[{\rm O}(4)\times {\rm O}(4)]$ manifold. As previously discussed, we are interested in constant solutions for all these functions. In order to satisfy the equations of motion, let us choose the non-vanishing components of $C^\mathcal{M}$ as follows: 
\begin{align}
    C^{0,\mathtt{I}=4}=C^{0}{}_{\mathtt{I}=4}= \epsilon\, c\,,
\end{align}
where $\epsilon=\pm1$. Below we present the values $\bar{\phi}^{i}$ of the non-vanishing  constant scalar fields, together with a suitable relation between the parameters $\Lambda$ and $L$, for each of the AdS$_3$ warpings presented in section \ref{sec2}.
\paragraph{Spacelike case:}
\begin{align}\label{solutionspace}
    \rho_{3}{}^{\mathtt{I}=4}=-\epsilon \rho_{3,\mathtt{I}=4}=\zeta\,,\qquad s_1&=-\xi-\chi\,, \qquad s_2=2s_1\,,\nonumber\\
    |c|=\frac{\sqrt{\lambda(1-\lambda)}}{\kappa \Lambda}\,,\qquad L&=-\epsilon\frac{\Lambda}{\sqrt{2\lambda}}\,, \qquad \chi=\frac13\left(\xi+\log\left(\pm\epsilon\frac{\sqrt\lambda}{\Lambda}\right)\right)\,.
\end{align}
\paragraph{Lightlike case:}
\begin{align}\label{solutionlight}
    \rho_{3}{}^{\mathtt{I}=4}=-\epsilon \rho_{3,\mathtt{I}=4}=\zeta\,,\qquad s_1&=-\xi-\chi\,, \qquad s_2=2s_1\,,\nonumber\\
    |c|=\frac{\sqrt{\alpha}\Lambda}{2\kappa}\,,\qquad L&=-\epsilon\frac{\Lambda}{\sqrt{2}}\,, \qquad \chi=\frac13(\xi-\log\Lambda)\,.
\end{align}
\paragraph{Timelike case:}
\begin{align}\label{solutiontime}
    \rho_{3}{}^{\mathtt{I}=4}=-\epsilon \rho_{3,\mathtt{I}=4}=\zeta\,,\qquad s_1&=-\xi-\chi\,, \qquad s_2=2s_1\,,\nonumber\\
    |c|=\frac{\sqrt{\lambda(\lambda-1)}}{\kappa \Lambda}\,,\qquad L&=-\epsilon\frac{\Lambda}{\sqrt{2\lambda}}\,, \qquad \chi=\frac13\left(\xi+\log\left(\pm\epsilon\frac{\sqrt\lambda}{\Lambda}\right)\right)\,.
\end{align}
These solutions correspond to backgrounds which uplift to six-dimensional spacetimes of the form ${\rm WAdS}_{3}\times M^3_{\xi,\zeta}$, $M^3_{\xi,\zeta}$ being a compact three-dimensional manifold with the topology of $\rm{S}^{3}$, whose metric depends on the vacuum values of the scalar fields $\xi$ and $\zeta$.\footnote{Mixings between the two factors would generate Kaluza-Klein vectors $A^\mathcal M$ in 3d transforming under SO$(4)$. However, this is not the case for the solutions under consideration, since we are choosing vectors which are singlets under that group.} For certain values of these scalars, as discussed in \cite{Eloy:2021fhc}, such internal space may reduce to a round 3-sphere or to a warped 3-sphere. In particular, we have
\begin{align}\label{loci}
    \mathrm{S}^3:&\qquad \zeta=0,\qquad \chi+\xi=0\,,\nonumber\\
    \mathrm{WS}^3:&\qquad \zeta=\sqrt{1-e^{-\Xi}},\qquad \chi+\xi=-\frac{\Xi}{6}\,.
\end{align}
For the locus defining the warped sphere to be well-defined, we must require $\Xi\geq0$, where the equality corresponds to the round sphere case. 
In the WS$^3$ locus, the above solutions feature two independent parameters, $\lambda$ and $\Xi$ and the non-vanishing field components correspond to those of \cite{Maurelli:2025iba}, some of which were found to be non-supersymmetric, for generic values of their parameters. For these reasons, while not addressing the precise uplift to type IIB supergravity of the above solutions, we are interested in assessing their stability.

Let us finally discuss the role of the $\epsilon$ sign appearing in the above solutions: in the WS$^3$ case, it defines the SU$(2)$ factor in the isometry group of S$^3$ which survives the warping, i.e. the ``chirality'' of the warped sphere. The gauge generators corresponding to the two SU$(2)$ factors in $\rm{SO}(4)_{gauge}\subset \mathcal{G}_{\mathrm{gauge}}$ read
\begin{equation}
   \mathcal{J}^m_{(L/R)}=(X^m{}_0+\Lambda\,\epsilon^{mnp}\,X_{n,p} )+\epsilon\,(X_{m,0}+\Lambda\,\epsilon^{mnp}\,X_{n}{}^p )\,,\,\,\,\,\epsilon=\pm 1\,.
\end{equation}
On the other hand, the chirality of the warping of AdS$_3$ depends on the sign of $L$. The two backgrounds given above, defined by different choices of $\epsilon$, correspond, in the locus, to different relative chiralities of WAdS$_3$ and WS$^3$. It is known that the supersymmetry of the solutions may depend on this feature \cite{El-Showk:2011euy,Maurelli:2026dpp}.

\section{Mass spectra and stability analysis}\label{sec6}
We are now ready to apply the general analysis of section \ref{sec3} to this model. For computational convenience, we shall consider a consistent truncation of the model described in the previous section, that retains the solutions \eqref{solutionspace}, \eqref{solutionlight}, and \eqref{solutiontime}. In order to do so, one can restrict to the singlet sector of a suitable invariance group of the gauge algebra. Let us recall that, in our model, the SO$(4,4)$ factor in the subgroup ${\rm SO}(3,3)\times {\rm SO}(1,1)\times {\rm SO}(4,4)$ only acts on the central charges in $T^8$, and its action stabilizes the embedding tensor. The solutions shown above survive the restriction to the singlet sector with respect to ${\rm SO}(3)\times {\rm SO}(3)\subset \mathrm{SO}(4,4)$. This is effectively implemented by setting $V_{\mathtt{I}=1,2,3}=V^{\mathtt{I}=1,2,3}=0$ in \eqref{funddec}. The mass spectrum in this consistent truncation will only differ from that of the original model in the degeneracy of certain mass eigenvalues and thus this choice will not affect the stability analysis.\footnote{In other words, every eigenspace of the mass spectrum contains a ${\rm SO}(3)\times {\rm SO}(3)$-singlet, which thus belongs to the truncation.} 
This truncation to the singlets under ${\rm SO}(3)\times {\rm SO}(3)$ is now defined by the scalar manifold:
\begin{equation}
    \mathcal{M}_{{\rm scal}}^{{\rm (trunc)}}=\frac{{\rm SO}(5,5)}{{\rm S}[{\rm O}(5)\times {\rm O}(5)]}\,,
\end{equation}
spanned by 25 scalar fields. Moreover, this restriction reduces the number of central charges from 8 to 2, without affecting the other gauge generators. Therefore, the gauge group of the truncated theory has dimension 14. 
Now the indices $\mathcal{M,N}$ label the adjoint representation of ${\rm SO}(5,5)$, which is 45-dimensional.
Accordingly, the embedding tensor $\Theta_{\mathcal M\mathcal N}$ is a $45\times45$ matrix with 
\begin{align}
    \mathrm{rank}(\Theta_{\mathcal M\mathcal N})=14.
\end{align}
A useful way of identifying the actual vectors participating in the minimal couplings, without explicitly breaking the global covariance of the model, is to use the \emph{rank factorization}  of the embedding tensor \cite{Trigiante:2016mnt,DallAgata:2023ahj}, that is to write it in the following way\footnote{The two matrices $\vartheta_{\mathcal M}{}^I$ and $\vartheta_{I\mathcal M}$ are not related by transposition, in general.}
\begin{align}
    \Theta_{\mathcal M\mathcal N}=\vartheta_{\mathcal M}{}^I\vartheta_{I\mathcal N}\,,\qquad I=1,\ldots,14\,.
\end{align}
We can define here the ``inverse'' matrix $\tilde \vartheta^{\mathcal MI}$ in such a way that $\vartheta_{I\mathcal M}\tilde \vartheta^{\mathcal MJ}=\delta_I^J$. The same contraction with $\vartheta_{\mathcal M}{}^I$ does not yield a Kronecker delta, but it provides a symmetric matrix $\theta^{IJ}=\tilde \vartheta^{\mathcal MI}\vartheta_{\mathcal M}{}^J$.
Finally, notice that such a decomposition is defined up to a $\mathrm{GL}(14)$ transformation of the indices $I,J$:
\begin{align}
    \vartheta'_{I\mathcal M}=R_I{}^J\vartheta_{J\mathcal M}\,,\qquad \vartheta'_{\mathcal M}{}^I=R^{-1}{}_J{}^I\,\vartheta_{\mathcal M}{}^J\,.
\end{align}
Using this freedom, one can always choose a frame in which the combination $C^\mathcal M\vartheta_\mathcal M{}^I$, which will appear in the expressions below, defines a single direction. Let us therefore split the index $I$ as $I=(U,\bullet)$, $\bullet$ denoting the direction singled out by $C^\mathcal M$. The previous rank-factorization further reduces to 
\begin{align}
    \Theta_{\mathcal M \mathcal N}&=\vartheta_{\mathcal M}{}^I\vartheta_{I\mathcal N}=\vartheta_{\mathcal M}{}^U\vartheta_{U\mathcal N}+\vartheta_{\mathcal M}{}^\bullet\vartheta_{\bullet\mathcal N}, \qquad C^\mathcal M\vartheta_{\mathcal M}{}^U=0\,.
\end{align}
The gauge Killing vectors can be split accordingly 
\begin{align}
    K_{\mathcal M}^i&=\Theta_{\mathcal M \mathcal N}\tilde K^{\mathcal N i}\equiv \vartheta_\mathcal M{}^IK_I^i=\vartheta_{\mathcal M}{}^U K_U^i+\vartheta_{\mathcal M}{}^\bullet K_\bullet^i, \qquad K_I^i= \vartheta_{I\mathcal M}\tilde K^{\mathcal Mi}\,.
\end{align}
On the backgrounds constructed above, and depending on the chosen values of $\xi$ and $\zeta$, the Killing vectors $\bar K_I^i=\{\bar K_U^i,\bar K_\bullet^i\}$ are not all linearly independent. In particular, since the manifold $M^3_{\xi,\zeta}$ has two isometries, two gauge Killing vectors among $\bar K_U^i$ always vanish on the background, thus leaving $12$ independent ones. In the locus $\zeta=\sqrt{1-e^{-\Xi}}$, corresponding to the ${\rm WS}^3$ space, two more gauge Killing vectors become zero, as the isometry group of the internal space is enhanced to $\mathrm{U}(1) \times \mathrm{SU}(2)_{L/R}$. Finally, in the case $\zeta=\Xi=0$, the vanishing gauge Killing vectors are $6$, leaving only $8$ of them linearly independent.

By making use of the aforementioned freedom in the choice of basis for the rank factorization, we can always rotate these vectors in such a way that the non-zero ones correspond to the first entries. This basis choice is therefore characterized by the splitting $I=(\hat I,\tilde I)=( u,\bullet\,,\tilde I)$, with $\bar K_{\tilde I}^i=0$ and $u$ labelling the non-zero vacuum gauge Killing vectors among $\bar K_U^i$. The vacuum Killing vectors $\bar K_{\hat I}^i$ can always be completed to a basis of the 25-dimensional space $T_{\bar\phi}\mathcal M_{\mathrm{scal}}$, by adding $25-\mathrm{rank}(\bar K_{\hat I}^i)$ orthonormal vectors $\bar\zeta_a^i$, characterized by the following properties:
\begin{align}
     0&=\bar K_u^i \bar{ \mathcal{G}}_{ij}\bar K_\bullet^j=\bar K_I^i \bar{\mathcal{G}}_{ij}\bar \zeta_a^j\,,\qquad
    \delta_{ab}=\bar \zeta_a^i\bar{\mathcal G}_{ij}\bar\zeta_b^j\,,\label{ortho}
\end{align}
where $\bar{\mathcal{G}}_{ij}$ is the metric of the scalar manifold on the background. This allows us to decompose the vector and scalar perturbations as 
\begin{align}
    \delta A_\mu ^\mathcal M\bar K_\mathcal M^i&=\delta A_\mu ^\mathcal M(\vartheta_{\mathcal M}{}^u\bar K_u^i+\vartheta_{\mathcal M}{}^\bullet \bar K_\bullet^i)\equiv \delta A_\mu ^u\bar K_u^i+\delta A_\mu ^\bullet \bar K_\bullet^i\,,\\
    \delta \phi^i&=\bar K_u^i\delta\phi^u+\bar K_\bullet^i\delta \phi^\bullet+\bar \zeta_a^i\delta\phi^a\,.
\end{align}
As we shall see below, the fluctuations $\delta\phi^{\hat I}=(\delta\phi^u,\,\delta\phi^\bullet)$ correspond to Goldstone bosons, while $\delta\phi^a$ represent the physical scalar fluctuations.

\subsection{Analysis of the perturbed equations}
In the following, we shall make the explicit derivation of the perturbed equations. To this end, let us collect here the following definitions
\begin{align}
    \bar{K}_{\mathcal M \mathcal N}&\equiv \bar{K}_\mathcal{M}^i\,\bar{G}_{ij}\, \bar{K}_\mathcal{N}^j=\vartheta_\mathcal M{}^I\bar{K}_{IJ}\vartheta_\mathcal N{}^J=\vartheta_\mathcal M{}^u\bar{K}_{uv}\vartheta_{\mathcal N}{}^v+\vartheta_\mathcal M{}^\bullet \bar{K}_{\bullet \bullet}\vartheta_{\mathcal N}{}^\bullet\,,\\
    \delta \hat A^{\hat{I}}_\mu&\equiv\delta  A^{\hat{I}}_\mu+\partial_\mu \delta \phi^{\hat{I}} \,,\label{massiveAvar}\\
    \mathcal M_{ij}&\equiv\frac12\overline{\partial_{ij}( |A|^2 K_{\bullet\bullet}+2V)}\,,\\
    \vartheta&\equiv C^\mathcal M\vartheta_\mathcal M{}^\bullet\,,\,\,\,\theta^{IJ}\equiv\tilde \vartheta^{\mathcal MI}\vartheta_{\mathcal M}{}^I\,,
\end{align}
and useful properties
\begin{align}
    0&=\Theta_{\mathcal M\mathcal N}X_{\mathcal P\mathcal Q}{}^\mathcal NC^\mathcal P\,,\label{prop1}\\
    0&=\bar K_I^k\overline{\partial_k K_{\bullet \bullet}}=\bar K_I^k\overline{\partial_k K_{u \bullet}}\,,\label{prop2}\\
    0&=\bar K_I^k\overline{\partial_k V}\,,\label{prop2.5}\\
    0&=\bar K_I^i \mathcal M_{ij}\,,\label{prop3}\\
    \bar K_I^i \overline{D_i K_{\bullet j}}&=-\frac12 \overline{\partial_j K_{I\bullet}}\,,\label{prop4}\\
\bar\zeta_a^i\overline{D_iK_{\bullet j}}\bar K^j_I&=\frac12\bar\zeta^i_a\overline{\partial_iK_{I\bullet}}\,, \label{prop5}\\
    0&={\bar\zeta}^i_{[b} {\overline{D_iK_{\bullet j}}} \bar{\zeta}^j_{a]}\,.\label{prop6}
\end{align}
Moreover, we shall fix the reparametrization gauge for the metric fluctuations $\delta h_{\mu\nu}$ in such a way that
\begin{equation}
    \bar{A}^\mu\,h_{\mu\nu}=0\,.\label{hgaugechoice}
\end{equation} 
See Appendix \ref{RGF} for a detailed derivation.

\paragraph{The Einstein Equations}\ \\
The linearized Einstein equations \eqref{EEq pert} can be rewritten as 
\begin{align}
    R^{(1)}_{\mu\nu}-V_0 h_{\mu\nu}-\vartheta K_{\bullet \hat I}\bar A_{(\mu}\delta \hat A^{\hat I}_{\nu)}-\left(\overline{\partial_iV }g_{\mu\nu}+\frac12\overline{\partial_i K_{\mathcal M\mathcal N}}\bar A^{\mathcal M}_{(\mu}\bar A^{\mathcal N}_{\nu)}\right)\delta \phi^i=0\,.
\end{align}
Let us emphasize here that, by virtue of \eqref{prop2} and \eqref{prop2.5}, the scalar fluctuations $\delta\phi^{\hat{I}}$ always enter the perturbed equations through the specific combination \eqref{massiveAvar}. This follows from the Higgs mechanism, by means of which the fluctuations $\delta \hat{A}^{\hat I}_{\mu}$ acquire a mass by combining with the corresponding Goldstone modes $\delta\phi^{\hat I}$, through a gauge transformation.

One can then contract the above equation with $\bar A^\mu \bar A^\nu$: by using \eqref{eq03} and the gauge fixing condition \eqref{hgaugechoice}, we obtain
\begin{align}\label{hexpr}
    \frac{m^2}{2}\left(|A|^2-\frac{(\bar A^\mu {\nabla}_\mu)^2}{m^2}\right)h-\vartheta K_{\bullet \hat I}|A|^2 a^{\hat I}=0\,,
\end{align}
where we defined $a^{\hat I}=\bar A^\mu \delta \hat A^{\hat I}_\mu$.

\paragraph{The Vector Equations}\ \\
The equation for the vector perturbations, in the chosen gauge, reads
\begin{align}
    \frac{2}{e}\epsilon^{\mu\nu\rho}\nabla_\nu\delta \hat A^I_\rho\vartheta_{I\mathcal M}+\vartheta_\mathcal M{}^I\bar K_{IJ}\delta\hat A^{J\mu}+\vartheta \bar A^\mu\left(\frac{h}{2}\bar K_{\bullet \mathcal M}+\bar \zeta_a^k\overline{\partial_k K_{\bullet \mathcal M}}\delta\phi^a\right)=0\,.
\end{align}
By contracting it with $\tilde \vartheta^{\mathcal MI}$ and by restricting, whenever possible, $I\to \hat I$, we get
\begin{align}\label{originalveceq}
     \frac{2}{e}\epsilon^{\mu\nu\rho}\nabla_\nu\delta \hat A^I_\rho+\theta^{I\hat J}\bar K_{\hat J\hat K}\delta \hat A^{\hat K\mu}+\vartheta\theta^{I\hat J}\bar A^\mu\left(\frac{h}{2}\bar K_{\bullet \hat J}+\bar \zeta_a^k\overline{\partial_k K_{\bullet \hat J}}\delta\phi^a\right)=0\,.
\end{align}
The above equation, depending on the free index $I=(\hat I,\tilde I)$, yields
\begin{align}\label{splitveceq}
    &\frac{2}{e}\epsilon^{\mu\nu\rho}\nabla_\nu\delta \hat A^{\hat I}_\rho+\Xi^{\hat I}{}_{\hat K}M^{\hat K\mu}=0\,,\qquad \frac{2}{e}\epsilon^{\mu\nu\rho}\nabla_\nu\delta  A^{\tilde I}_\rho+\Xi^{\tilde I}{}_{\hat K} M^{\hat K \mu}=0\,,
\end{align}
where we defined 
\begin{align}\label{defXi}
    \Xi^I{}_{\hat J}=\theta^{I\hat K}\bar K_{\hat K\hat J}\,,\qquad M^{\hat I \mu}=\delta \hat A^{\hat I\mu}+\vartheta \bar{A}^{\mu}(K^{-1}){}^{\hat{I}\hat{J}}\left(\frac{h}{2}\bar K_{\bullet \hat J}+\bar \zeta_a^k\overline{\partial_k K_{\bullet \hat J}}\delta\phi^a\right)\,.
\end{align} 
One can prove that the combination 
\begin{align}
    \delta\mathcal A^{\hat I}_\mu= \Upsilon^{\hat I}{}_{\tilde I}\delta A^{\tilde I}_\mu + P^{\hat I}{}_{\hat J}(\Xi^{-1})^{\hat J}{}_{\hat K}\delta \hat A^{\hat K}_\mu\,,\qquad \Xi^{\tilde I}{}_{\hat I}\Upsilon^{\hat I}{}_{\tilde J}=\delta^{\tilde I}_{\tilde J}\,, \qquad \Upsilon^{\hat I}{}_{\tilde I} \Xi^{\tilde I}{}_{\hat J}=P^{\hat I}{}_{\hat J}\,,\qquad  P^2=P
\end{align}
is massless, as it satisfies
\begin{align}
    \frac{2}{e}\epsilon^{\mu\nu\rho}\nabla_\nu\delta \mathcal A^{\hat I}_{\rho}=0\,.
\end{align}
These are the $14-\mathrm{rank}(\bar K_{\hat I}^i)$ vectors that are not involved in the Higgs mechanism. It is therefore natural to consider $\delta A^I=(\delta \hat A^{\hat I}, \delta\mathcal A^{\hat I})$ as the independent vector fluctuations.

\paragraph{The Scalar Field Equations}\ \\
Using the definitions given earlier, the equations of motion for the scalar perturbations can be written in the form
\begin{align}
    0&=\vartheta\bar{\mathcal G}_{ij} \bar K_\bullet^j\Big(-\nabla^\mu h_{\mu\nu}\bar A^\nu+\frac12 \nabla^\mu h \bar A_\mu\Big)+\bar{\mathcal G}_{ij}\bar K_I^j\nabla^\mu \delta\hat A_\mu^I+\bar{\mathcal G}_{ij}\nabla^\mu \partial_\mu \delta\phi^a\bar\zeta_a^j-\mathcal M_{ij}\bar\zeta_a^j\delta\phi^a\nonumber\\
    &+2\vartheta \overline{D_i K_\bullet^m}\bar{\mathcal G}_{mn}\bar A^\mu\left(\frac\vartheta2 \bar K_\bullet^n h_{\mu\nu}\bar A^\nu-\bar K_I^n\delta\hat A^I_\mu\right)-2\vartheta \overline{D_i K_\bullet^m}\bar{\mathcal G}_{mn}\bar A^\mu\partial_\mu \delta\phi^a\bar\zeta_a^n\,.
\end{align}
Using the properties \eqref{ortho} and \eqref{prop1}-\eqref{prop5}, one can now consider the different projections along the index $i$: the projections along  $\bar K_\bullet^i$ and $\bar K_u^i$ are identically satisfied. As for the 
projection along $\bar \zeta_b^i$ we find: 
\begin{align}
    0&=(\bar \zeta^i_a\bar{\mathcal G}_{ij}\bar\zeta_b^j)\nabla^\mu \partial_\mu \delta\phi^a-(\bar \zeta_a^i\mathcal M_{ij}\bar\zeta_b^j)\delta\phi^a-2\vartheta \bar \zeta^i_{[b}\overline{D_i K_{\bullet j}}\bar\zeta_{a]}^j\bar A^\mu\partial_\mu \delta\phi^a\nonumber\\
    &+\vartheta \bar A^\mu\left(\frac\vartheta2 (\bar\zeta^i_b\partial_i K_{\bullet\bullet}) h_{\mu\nu}\bar A^\nu-(\bar\zeta^i_b\partial_i K_{I\bullet})\delta\hat A^I_\mu\right)\,,
\end{align}
where we used \eqref{divergenceeq}. If we now use that $\bar\zeta_a^i$ are orthonormal, the gauge fixing condition, and \eqref{prop6}, we find
\begin{align}\label{finalphia}
    0&=\nabla^\mu \partial_\mu \delta\phi^a-\mathcal M^a{}_b\delta\phi^b-\vartheta (\bar\zeta^{ai}\partial_i K_{\hat I\bullet})a^{\hat I}\,,
\end{align}
where we defined $\mathcal M_{ab}=\bar \zeta_a^i\mathcal M_{ij}\bar\zeta_b^j$. Let us observe that the scalar combination $a^{\hat I}$ provides a source term for the scalar fields $\delta\phi^a$, thus we will now derive the Klein--Gordon equation satisfied by such scalar combination. To this end, let us act with the differential operator $\epsilon_{\mu\alpha\beta}\bar A^\alpha\nabla^\beta$ on the first equation in \eqref{splitveceq}, yielding
\begin{align}
        \square a^{\hat I}-\left(\frac m2\Xi^{\hat I}{}_{\hat J}+\frac14 \Xi^{\hat I}{}_{\hat K}\Xi^{\hat K}{}_{\hat J}\right)a^{\hat J}-(\bar A^\nu \nabla_\nu)(\nabla_\beta \delta \hat A^{\hat I\beta})-\frac\vartheta4 |A|^2\Xi^{\hat I}{}_{\hat K}\theta^{\hat K\hat J}\left(\frac{h}{2}\bar K_{\bullet \hat J}+\bar \zeta_a^k\overline{\partial_k K_{\bullet \hat J}}\delta\phi^a\right)=0\,,\nonumber
\end{align}
where the tensor $\Xi^{\hat I}{}_{\hat J}$ was defined in \eqref{defXi}. Moreover, we used 
\begin{align}\label{divergenceeq}
        \bar K_{\hat I\hat J}\nabla_\mu \delta \hat A^{\hat J\mu}+\vartheta\bar A^\mu\left(\frac12\nabla_\mu h\bar K_{\bullet \hat I}+\bar \zeta_a^k\overline{\partial_k K_{\bullet \hat I}}\nabla_\mu\delta\phi^a\right)=0\,,
\end{align}
obtained by contracting \eqref{splitveceq} with $\nabla_\mu$. This allows us to further simplify the expression as follows:
\begin{align}
        &\square a^{\hat I}-\left(\frac m2\Xi^{\hat I}{}_{\hat J}+\frac14 \Xi^{\hat I}{}_{\hat K}\Xi^{\hat K}{}_{\hat J}\right)a^{\hat J}\nonumber\\
        &-\frac\vartheta4 \left(|A|^2\Xi^{\hat I}{}_{\hat K}-4(\Xi^{-1})^{\hat I}{}_{\hat K}(\bar A^\rho \nabla_\rho)^2\right)\theta^{\hat K\hat J}\left(\frac{h}{2}\bar K_{\bullet \hat J}+\bar \zeta_a^k\overline{\partial_k K_{\bullet \hat J}}\delta\phi^a\right)=0\,,
\end{align}
which has the following structure 
\begin{align}\label{finalaI}
    \square a^{\hat I}-M^{\hat I}{}_{\hat J}a^{\hat J}-N^{\hat I}{}_a\delta \phi^a=0\,.
\end{align}

In order to diagonalize the mass matrix for the whole set of scalar field fluctuations $(a^{\hat I},\delta \phi^a)$

\begin{align}\label{Big M}
    M=\begin{pmatrix}
    M^{\hat I}{}_{\hat J} && N^{\hat I}{}_a\\
    \vartheta (\bar\zeta^{ai}\partial_i K_{\hat I\bullet}) && \mathcal M_{ab}
    \end{pmatrix}\,,
\end{align}
it is necessary to derive the expression of $h$ from \eqref{hexpr}. This is easily achieved by Fourier expanding all fields, as done in section \ref{sec3}. The resulting system of differential equations then reduces to a set of decoupled Schr\"odinger equations, each of the form
\begin{align}\label{genericschroed}
    -\Psi''_X+V(\nu_X;p,\omega)\Psi_X=-\mathcal E_{\mathrm{Sch}}(p,\omega)\Psi_X\,, \qquad X=1,\ldots,25\,,
\end{align}
where $\Psi_X=\Psi_X(p,\omega)$ denotes the radial wavefunction associated with a generic Fourier mode. This reduction to a Schr\"odinger problem allows us to apply the stability criteria developed in section \ref{sec3}. Due to the mixing between $\delta\phi^a$ and $a^{\hat I}$, the masses $m_X$ appearing in \eqref{genericschroed} depend non-trivially on the Fourier momenta and frequencies, as anticipated. Below we specify the resulting equations and the corresponding scalar field mass spectrum for each AdS warping.

\subsection{Spectra}
\paragraph{Spacelike case:} \ \\
Since $A^\mu \partial_\mu =\frac{\kappa}{\lambda} \partial_x$, we can consider the following Fourier expansion for a generic scalar field
\begin{align}
    \boldsymbol{\Phi} = \int\frac{\dd p}{2\pi}e^{\ri p x} \boldsymbol{\varphi}(t,z)\,,
\end{align}
$\boldsymbol{\Phi}$ denoting $h$, $a^{\hat I}$ or $\delta \phi^a$. In this way, the explicit perturbed equations become
\begin{align}
    h&=\frac{2\lambda\vartheta}{m^2\lambda +p^2}\bar K_{\bullet \hat I}a^{\hat I}\,,\nonumber\\
    0&=\square a^{\hat I}-\frac12\left(m\Xi^{\hat I}{}_{\hat J}+\frac12 \Xi^{\hat I}{}_{\hat K}\Xi^{\hat K}{}_{\hat J}+\frac{\vartheta^2}{2}\frac{\kappa^2}{m^2\lambda+p^2}\left(\Xi^{\hat I}{}_{\hat K}+\frac{4p^2}{\lambda}(\Xi^{-1})^{\hat I}{}_{\hat K}\right)\theta^{\hat K\hat L}K_{\bullet\hat L}K_{\bullet \hat J}\right)a^{\hat J}\nonumber\\
    &-\frac{\vartheta\kappa^2}{4\lambda} \left(\Xi^{\hat I}{}_{\hat K}+\frac{4p^2}{\lambda}(\Xi^{-1})^{\hat I}{}_{\hat K}\right)\theta^{\hat K\hat L}\bar \zeta_a^k\overline{\partial_k K_{\bullet \hat L}}\delta\phi^a=0\,,\nonumber\\
    0&=\nabla^\mu \partial_\mu \delta\phi^a-\mathcal M^a{}_b\delta\phi^b-\vartheta (\bar\zeta^{ai}\partial_i K_{\hat I\bullet})a^{\hat I}\,.
\end{align}
The eigenvalues of the mass matrix \eqref{Big M}, together with their multiplicity, are 
\[
\begin{array}{c@{\qquad\qquad}c}

\begin{array}[t]{c|l}
\multicolumn{2}{c}{\mathrm{WS}^3} \\[0.5ex]
\hline
5  & 0 \\[0.5ex]
10 & \displaystyle
\frac{4}{L^2}\left(e^{\Xi}-1\right)
\\[1ex]
5  & \displaystyle
\frac{4}{L^2}
-\frac{4}{L^2}
\sqrt{1+L^2p^2\left(\frac1\lambda-1\right)}
\\[1ex]
5  & \displaystyle
\frac{4}{L^2}
+\frac{4}{L^2}
\sqrt{1+L^2p^2\left(\frac1\lambda-1\right)}
\end{array}

&

\begin{array}[t]{c|l}
\multicolumn{2}{c}{\mathrm{S}^3} \\[0.5ex]
\hline
15 & 0 \\[0.5ex]
5  & \displaystyle
\frac{4}{L^2}
-\frac{4}{L^2}
\sqrt{1+L^2p^2\left(\frac1\lambda-1\right)}
\\[1ex]
5  & \displaystyle
\frac{4}{L^2}
+\frac{4}{L^2}
\sqrt{1+L^2p^2\left(\frac1\lambda-1\right)}
\end{array}

\end{array}
\]
Observe that for any $\Xi>0$, $0<\lambda\leq 1$ and $p\in\mathbb{R}$, these masses lead to real values of \eqref{nuspacelike}.
Indeed, the expression of $\nu^2$ for the momentum-dependent masses simplifies to a perfect square
\begin{align}\label{perfectsquare}
    \nu^2=\left(1\pm\frac12\sqrt{ 1 +L^2 p^2\left(\frac1\lambda-1\right) }\right)^2\geq0\,.
\end{align}
The above result in the $\lambda=1$ case, corresponding to round AdS$_3$, matches the one in \cite{Eloy:2021fhc}.
\paragraph{Lightlike case:} \ \\
Following the discussion for the spacelike case, the perturbed equations for the lightlike warping read
\begin{align}
    &(\bar A^\mu \nabla_\mu)^2h=0\,,\nonumber\\
    &\square a^{\hat I}-\frac12\left(m\Xi^{\hat I}{}_{\hat J}+\frac12 \Xi^{\hat I}{}_{\hat K}\Xi^{\hat K}{}_{\hat J}\right)a^{\hat J}-\vartheta\frac{4\kappa^2p^2}{L^4} (\Xi^{-1})^{\hat I}{}_{\hat K}\theta^{\hat K\hat L}\bar \zeta_a^k\overline{\partial_k K_{\bullet \hat L}}\delta\phi^a=0\,,\nonumber\\
    &\nabla^\mu \partial_\mu \delta\phi^a-\mathcal M^a{}_b\delta\phi^b-\vartheta (\bar\zeta^{ai}\partial_i K_{\hat I\bullet})a^{\hat I}=0\,.
\end{align}
By diagonalizing the corresponding total mass matrix, we get the following masses and multiplicities: 
\[
\begin{array}{c@{\qquad\qquad}c}

\begin{array}[t]{c|l}
\multicolumn{2}{c}{\mathrm{WS}^3} \\[0.5ex]
\hline
5  & \displaystyle 0 \\[0.5ex]
10 & \displaystyle
\frac{4}{L^2}\left(e^{\Xi}-1\right)
\\[1ex]
5  & \displaystyle
\frac{4}{L^2}
\left(1-\sqrt{1+4\alpha L^2p^2}\right)
\\[1ex]
5  & \displaystyle
\frac{4}{L^2}
\left(1+\sqrt{1+4\alpha L^2p^2}\right)
\end{array}

&
\begin{array}[t]{c|l}
\multicolumn{2}{c}{\mathrm{S}^3} \\[0.5ex]
\hline
15 & \displaystyle 0 \\[0.5ex]
5  & \displaystyle
\frac{4}{L^2}
\left(1-\sqrt{1+4\alpha L^2p^2}\right)
\\[1ex]
5  & \displaystyle
\frac{4}{L^2}
\left(1+\sqrt{1+4\alpha L^2p^2}\right)
\end{array}

\end{array}
\]
Crucially, the parameter $\nu$ in \eqref{nulightlike} is real for any considered value of the parameters $\Xi$, $p$, $\alpha$. Indeed, notice that for the $p$-dependent masses, we get
\begin{align}
    \nu^2=\left(2\pm\sqrt{1+4 \alpha  L^2 p^2})\right)^2\geq0\,.
\end{align}
\paragraph{Timelike case:} \ \\
Let us finally address the timelike WAdS$_3$ case: since $A^\mu \partial_\mu =\frac{\kappa}{\lambda} \partial_t$, we can consider the following Fourier expansion for a generic scalar field
\begin{align}
    \boldsymbol{\Phi} = \int\frac{\dd \omega}{2\pi}e^{-\ri \omega t} \boldsymbol{\varphi}(x,z)\,.
\end{align}
The resulting equations for the scalar fluctuations in the timelike case are
\begin{align}\label{solh TL}
    h &=\frac{2\lambda\vartheta}{m^2\lambda -\omega^2}\bar K_{\bullet \hat I}a^{\hat I}\,,\nonumber\\
    0&=\square a^{\hat I}-\frac12\left(m\Xi^{\hat I}{}_{\hat J}+\frac12 \Xi^{\hat I}{}_{\hat K}\Xi^{\hat K}{}_{\hat J}+\frac{\vartheta^2}{2}\frac{\kappa^2}{m^2\lambda-\omega^2}\left(-\Xi^{\hat I}{}_{\hat K}+\frac{4\omega^2}{\lambda}(\Xi^{-1})^{\hat I}{}_{\hat K}\right)\theta^{\hat K\hat L}K_{\bullet\hat L}K_{\bullet \hat J}\right)a^{\hat J}\nonumber\\
    &-\frac{\vartheta\kappa^2}{4\lambda} \left(- \Xi^{\hat I}{}_{\hat K}+\frac{4\omega^2}{\lambda}(\Xi^{-1})^{\hat I}{}_{\hat K}\right)\theta^{\hat K\hat L}\bar \zeta_a^k\overline{\partial_k K_{\bullet \hat L}}\delta\phi^a\,,\nonumber\\
    0&=\nabla^\mu \partial_\mu \delta\phi^a-\mathcal M^a{}_b\delta\phi^b-\vartheta (\bar\zeta^{ai}\partial_i K_{\hat I\bullet})a^{\hat I}\,,
\end{align}
while the corresponding masses and multiplicities are shown in the table below:
\[
\begin{array}{c@{\qquad\qquad}c}

\begin{array}[t]{c|l}
\multicolumn{2}{c}{\mathrm{WS}^3} \\[0.5ex]
\hline
5  & \displaystyle 0 \\[0.5ex]
10 & \displaystyle
\frac{4}{L^2}\left(e^{\Xi}-1\right)
\\[1ex]
5  & \displaystyle
\frac{4}{L^2}
-\frac{4}{L^2}
\sqrt{1+L^2\omega^2\left(1-\frac1\lambda\right)}
\\[1ex]
5  & \displaystyle
\frac{4}{L^2}
+\frac{4}{L^2}
\sqrt{1+L^2\omega^2\left(1-\frac1\lambda\right)}
\end{array}

&
\begin{array}[t]{c|l}
\multicolumn{2}{c}{\mathrm{S}^3} \\[0.5ex]
\hline
15 & \displaystyle 0 \\[0.5ex]
5  & \displaystyle
\frac{4}{L^2}
-\frac{4}{L^2}
\sqrt{1+L^2\omega^2\left(1-\frac1\lambda\right)}
\\[1ex]
5  & \displaystyle
\frac{4}{L^2}
+\frac{4}{L^2}
\sqrt{1+L^2\omega^2\left(1-\frac1\lambda\right)}
\end{array}

\end{array}
\]
For the allowed values of the parameters, that is for $\Xi>0$, $\lambda\geq1$ and $\omega\in \mathbb R$, all masses correspond to positive values of $\nu^2$, see \eqref{nutimelike}. As for the spacelike and lightlike warpings, the $\omega$-dependent masses form perfect squares, as in \eqref{perfectsquare}.

\section{Conclusions}\label{sec7}

In this work, we investigate the perturbative stability of warped anti-de Sitter spacetimes in three-dimensional gauged supergravity. Our analysis proceeds in two complementary steps. First, we study the stability of a massive Klein--Gordon scalar field propagating on WAdS$_3$ backgrounds and identify criteria for establishing perturbative stability. The analysis is based on reducing the Klein--Gordon equations to a one-dimensional Schr\"odinger problem. For the different WAdS$_3$ geometries, the resulting stability condition can be expressed in terms of the reality of a parameter $\nu$, which depends on the mass of the scalar fluctuation and on its Fourier modes.

Secondly, we embed the warped geometries into the half-maximal $D=3$ gauged supergravity arising from the compactification of $D=6$, $\mathcal{N}=(1,1)$ supergravity on a three-sphere. A key insight in this procedure is the physical interpretation of the warping parameters. Indeed, while non-vanishing scalar fields in the three-dimensional theory affect the uplifted background through the deformation of the internal space (as explicitly shown in \cite{Eloy:2021fhc}), the deformation of the non-compact part is instead induced by a background vector field. Specifically, the warpings discussed in the present paper require turning on vectors that gauge central charges of the three-dimensional gauge algebra. These vectors are proportional to the characteristic one-form that deforms $\mathrm{AdS}_3$. We then analyze the spectrum of physical scalar fluctuations around explicit WAdS$_3$ solutions, by disentangling the corresponding linearized equations.

Our results show that, for the $\rm{S}^3$ and $\rm{WS}^3$ loci \eqref{loci}, all mass eigenvalues lead to real values of the $\nu$ parameter, for each warping, thus ensuring the perturbative stability of the chosen backgrounds. This provides an explicit realisation of the stability criterion developed in section \ref{sec3} within three-dimensional gauged supergravity. An important feature of the spectra is that the scalar masses generated by the supergravity dynamics need not be momentum-independent; rather, this dependence is precisely what ensures the fulfillment of the stability criteria.

It would be interesting to extend the present results beyond three-dimensional supergravity to include the Kaluza--Klein modes of the higher-dimensional uplift. This would require an extension of the AdS$_3$ Kaluza-Klein mass formulae of \cite{Eloy:2020uix} to more general backgrounds along the lines of \cite{Eloy:2026jzg}. This analysis should also be combined with the study of the supersymmetry of these backgrounds, either directly in three dimensions or by verifying whether they precisely uplift to the Type IIB configurations constructed in \cite{Maurelli:2025iba}, whose supersymmetry properties are already known.

Finally, a natural direction for future work would be to generalize the techniques developed here to the non-supersymmetric deformations of $\mathrm{AdS}_5\times\mathrm{S}^5$, introduced in \cite{Maldacena:2008wh}, which are holographically dual to non-relativistic four-dimensional field theories. Both factors in the geometry are warped by means of $1$-forms, closely paralleling the solutions considered in the present paper. Such analysis would require developing the appropriate criteria to establish perturbative stability and to map these backgrounds to suitable solutions of SO$(6)$-gauged maximal five-dimensional supergravity.

\section*{Acknowledgements}
R.N. was supported by the European Union and the Czech Ministry of Education, Youth and Sports
(Project: MSCA Fellowship CZ FZU III - CZ.02.01.01/00/22 010/0008598). MO is supported by AEI-Spain PID2023-152148NB-I00 and by Maria de Maeztu excellence unit grant CEX2023-001318-M, by Xunta de Galicia (CIGUS Network of Research Centres and project ED431F-2023/19), and by the European Union FEDER.

\appendix

\section{Details of the computations in section \ref{sec3}}\label{app0}
In this appendix, we provide the explicit derivation of the main formulae in section \ref{sec3}. 

\subsection{AdS$_3$ case}
When discussing stability on AdS$_3$, we distinguished two main classes of solutions to the Schr\"odinger problem: $\omega\in \mathbb R$, $\nu^2>0$ and $\omega\in\ri \mathbb R$, $\nu^2<0$. 
\paragraph{The $\omega\in \mathbb R$, $\nu^2>0$ case.} As mentioned in subsection \ref{sec31}, this class of solutions had the form \eqref{schsolads}. Finiteness at $z=0$, implies $B_{\omega,p}=0$, while the orthogonality relation \eqref{orthogonality} is satisfied by virtue of \cite{Leon_2015}
\begin{align}\label{leon}
    \int_0^\infty \dd z z J_\nu (k z)J_{\nu}(k'z)= \frac{\delta(k-k')}{k}\,, \qquad \nu\geq -1, \qquad k,k'>0.
\end{align}
The above relation requires restricting to modes for which $\mathcal E_{\mathrm{Sch}}>0$, that is $\omega^2>p^2$. Equation \eqref{Energy1} is then derived by applying \eqref{leon} to \eqref{Energy0} as follows
\begin{align}
    E(t)&=\frac14\int \frac{\dd \omega\dd \omega' \dd p}{(2\pi)^3}e^{-\ri (\omega-\omega')t}(\omega+\omega')^2\frac{1}{\sqrt{\omega^2-p^2}}\delta(\sqrt{\mathcal{E}_{\mathrm{Sch}}}-\sqrt{\mathcal {E'}_{\mathrm{Sch}}})A_{\omega,p}A^\dagger_{\omega',p}\nonumber\\
    &=\frac14\int \frac{\dd \omega\dd \omega' \dd p}{(2\pi)^3}e^{-\ri (\omega-\omega')t}(\omega+\omega')^2\frac{1}{|\omega'|}\delta(\omega-\omega')A_{\omega,p}A^\dagger_{\omega',p}\nonumber\\
    &=\int \frac{\dd \omega \dd p}{(2\pi)^2}|\omega| |A_{\omega,p}|^2\,,
\end{align}
which is manifestly positive and time-independent. Finiteness is achieved by suitably combining the single mode solutions, that is by choosing the coefficients $A_{\omega,p}$. 
\paragraph{The $\omega\in\ri \mathbb R$, $\nu^2<0$ case.} The solution for this range of parameters has the form \eqref{Hankel1}, which is obtained by choosing $B_{\beta,p}=\ri A_{\beta,p}$. Such requirement ensures that the total solution
\begin{align}\label{alternativesol}
    \phi(t,x,z)=\int \frac{\dd \beta \dd p}{(2\pi)^2}e^{-\beta t+\ri p x}\sqrt{\frac zL}\,\psi_{\beta,p}(z)\,,
\end{align}
vanishes at $z=\infty$.
Given the above expansion, the energy reads
\begin{align}
    E(t)=-\frac14 \int_0^\infty \dd z\int\frac{\dd \beta\dd \beta' \dd p}{(2\pi)^3} e^{-(\beta+\beta')t}(\beta-\beta')^2\psi_{\beta,p}\psi^\dagger_{\beta',p}\,,
\end{align}
while the boundary conditions \eqref{boundarycond} are rewritten as
\begin{align}\label{hankelmagic}
    0=\left.(\psi_{\omega,p}\partial_z\psi^\dagger_{\omega',p}-\psi^\dagger _{\omega',p}\partial_z\psi_{\omega,p})\right\vert_0^\infty=A^\dagger_{\beta',p}A_{\beta,p}\frac{4}{\pi}(1+\coth\pi\gamma)\sin\left[ \frac\gamma2\log \left|\frac{\mathcal E'_{\mathrm{Sch}}}{\mathcal E_{\mathrm{Sch}}}\right|\right]\,,
\end{align}
which requires a quantization of the Schr\"odinger energy written in \eqref{quantisationE}.
The energy then takes the form
\begin{align}
    E(t)&=-\frac14\int\frac{\dd\beta\dd\beta'\dd p}{(2\pi)^3}\frac{e^{-(\beta+\beta') t}(\beta-\beta')^2}{|\mathcal E_{\mathrm{Sch}}'|-|\mathcal E_{\mathrm{Sch}}|}\frac{4}{\pi}(1+\coth\pi\gamma)\sin\left[  \frac\gamma2\log \left|\frac{\mathcal E'_{\mathrm{Sch}}}{\mathcal E_{\mathrm{Sch}}}\right|\right]A^\dagger_{\beta',p}A_{\beta,p}\,,
\end{align}
which shows that the only non-vanishing contributions to the energy arise when $\beta'=-\beta$. We restrict to states within the discrete spectrum \eqref{quantisationE} by choosing the Fourier coefficients as follows
\begin{align}
    A_{\beta,p}=2\pi\sum_{n=-\infty}^{\infty}\left(\delta(\beta-\beta_n)+\delta(\beta+\beta_n)\right)a_{\beta_n,p}\,,
\end{align}
where $\beta_n=\sqrt{\mathcal {E}_n-p^2}$.
From the products of the delta functions, only the mixed products with the same level $n$ contribute to the expression. The energy is then computed to be
\begin{align}
    E(t)&=-\frac{4\gamma}{\pi}(1+\coth\gamma\pi)\sum_{n=-\infty}^\infty\int_{-\sqrt {\mathcal E_n}}^{\sqrt {\mathcal E_n}}\frac{\dd p}{2\pi}\frac{\mathcal E_n-p^2}{E_n}|a_{\beta_n,p}|^2\,,
\end{align}
where the range of integration in $p$ is restricted by the reality of $\beta_n$.
If we assume, for instance, $a_{\beta_n,p}$ not to depend on $p$ we get
\begin{align}\label{Energy0final}
    E(t)&=-\frac{8\gamma}{3\pi^2}(1+\coth\pi\gamma)\sum_{n=-\infty}^\infty e^{\frac{\pi n}{\gamma}}\sqrt{\mathcal E_0}|a_{\beta_n}|^2\,,
\end{align}
which is manifestly negative and unbounded from below.

\subsection{WAdS$_3^+$ case}
The conservation of the energy is ensured by the fact that the following equation holds 
\begin{align}\label{orthowhittaker}
    \int\frac{\dd p}{2\pi}\left.\left(\psi_{\beta,p}\partial_z\psi^\dagger_{\beta',p}-\psi^\dagger_{\beta',p}\partial_z\psi_{\beta,p}\right)\right\vert^\infty_0=2\ri\int\frac{\dd p}{2\pi}\frac{e^{\frac{(m+n)\pi}{\gamma}}}{\cosh \pi \gamma}\beta_0\left(\sinh\frac{\pi}{2} Lp \right)A^\dagger_{\beta',p} A_{\beta,p}=0\,,
\end{align}
whenever $\beta,\beta'$ are in the spectrum \eqref{betaspectrum} and if $A^\dagger_{\beta',p} A_{\beta,p}$ is an even function of $p$. 

In order to show that the energy is negative and unbounded from below, let us first restrict $\beta,\beta'$ to their spectrum, by redefining
\begin{align}
    A_{\beta,p}=2\pi \sum_{n=-\infty}^\infty(\delta(\beta-\beta_n)+\delta(\beta+\beta_n))a_{\beta_n,p}\,.
\end{align}
Since the energy \eqref{energyspacelike2} is peaked around $\beta'=-\beta$, only the mixed product of delta functions contribute, yielding 
\begin{align}
    E(t)=-2 \sqrt{\lambda } \beta_0\sum_{n=-\infty}^{\infty}\int_{-p_0}^{p_0}\frac{\dd p}{2\pi}    \frac{\pi\gamma e^{\frac{2 \pi  n}{\gamma }}}{\sinh2\pi\gamma}  2 \mathrm{Re}\left(\frac{a_{\beta_n,p}^\dagger a_{-\beta_n,p}}{\Gamma \left(\frac{i L p}{2}-i \gamma +\frac{1}{2}\right) \Gamma \left(\frac{i L p}{2}+i \gamma +\frac{1}{2}\right)}\right)\,.
\end{align}
We can restrict to states with momentum in a suitable neighborhood of $p=0$ such that the integrand is positive. The energy is therefore negative and can be made arbitrarily large, in absolute value, by increasing the value of $\beta_0$.

\subsection{WAdS$_3^0$ case}
The conserved energy for the configuration \eqref{solutionWAdS0}, with $\mathcal{E}_{\mathrm{Sch}}<0$ and $\nu^2<0$ is
\begin{align}\nonumber
    E(x_-)&=-\frac{1}{2\pi}\int\frac{\dd \omega\dd\omega'\dd p}{(2\pi)^3}e^{-\ri (\omega-\omega')x_-}\frac{\omega'+\omega}{\omega'-\omega}(1+\coth \pi\gamma)\theta(-\omega p)\theta(-\omega'p)a_{\omega,p}a^{\dagger}_{\omega',p}\sin\frac{\gamma}{2}\log\frac{\omega'}{\omega}\,.
\end{align}
The integrand is only non-vanishing when $\omega'=\omega$: we can select the non-zero contributions by choosing 
\begin{align}
a_{\omega,p}=2\pi\sum_{n=-\infty}^\infty \delta(\omega-\omega_n)a_{\omega_n p},
\end{align} 
which yields
\begin{align}
    E(x_-)&=-\frac{1}{\pi}\sum_{n=-\infty}^\infty \int_{-p_0}^{p_0}\frac{\dd p}{2\pi}\theta(-4\omega_n p)\gamma(1+\coth\pi\gamma)|a_{\omega_n p}|^2\nonumber\\
    &=-\frac{1}{\pi}\sum_{n=-\infty}^\infty \int_{-p_0}^{0}\frac{\dd p}{2\pi}\gamma(1+\coth\pi\gamma)|a^+_{\omega_n p}|^2-\frac{1}{\pi}\sum \int_{0}^{p_0}\frac{\dd p}{2\pi}\gamma(1+\coth\pi\gamma)|a^-_{\omega_n p}|^2\nonumber\\
    &=-\frac{2}{\pi}\sum_{n=-\infty}^\infty \int_{0}^{p_0}\frac{\dd p}{2\pi}\gamma(1+\coth\pi\gamma)|a^-_{\omega_n p}|^2\,,
\end{align}
where  $a^\pm_{\omega_n p}=a_{\omega_n p}({\omega_n=\pm |\omega_n|})$  and $a^+_{\omega_n p}=(a^-_{\omega_n -p})^\dagger$, due to the reality condition. The above expression is clearly negative and can be made unbounded from below by suitably choosing $a^-_{\omega_n p}$.

\section{Reparametrization Gauge Choice}\label{RGF}
We now discuss the gauge-fixing condition introduced in \eqref{hgaugechoice}. Under an infinitesimal diffeomorphism generated by $\xi^\mu$, the metric fluctuation $h_{\mu\nu}$ transforms as follows:
$$\delta h_{\mu\nu}=2\bar{\nabla{}}_{(\mu}\xi_{\nu)}\,.$$
Using the explicit form of $\bar A^\mu$ for each type of warping, the gauge-fixing condition yields a system of differential equations for the components $\xi_\mu$. We give these equations below and discuss how they can be solved in the spacelike, lightlike, and timelike cases.

\paragraph{Spacelike case.}
Using the explicit expression for $\bar A^\mu$, the gauge-fixing condition gives
\begin{align}
    h'_{\mu x}=h_{\mu x}+2\nabla_{(\mu} \xi_{x)}\,,
\end{align}
which are equivalent to
\begin{align}
    0&=h'_{x x}=h_{x x}+2\nabla_{(x} \xi_{x)}=h_{x x}+2\partial_{x} \xi_{x}\,,\quad \implies \xi_x=-\frac12\int \dd x' h_{xx}(t,x',z)+ \zeta(t,z)\,,\nonumber\\
    0&=h'_{t x}=h_{t x}+2\nabla_{(t} \xi_{x)}=h_{t x}+\partial_{t} \xi_{x}+\partial_{x} \xi_{t}+2\frac{\lambda}{L}\xi_z\,,\quad \implies \xi_z=-\frac{L}{2\lambda}(h_{t x}+\partial_{t} \xi_{x}+\partial_{x} \xi_{t})\nonumber\\
    0&=h'_{z x}=h_{z x}+2\nabla_{(z} \xi_{x)}=h_{z x}+\partial_{z} \xi_{x}+\partial_{x} \xi_{z}+2\frac{\lambda}{L}\xi_t+\frac{\lambda}{z}\xi_x\,,\quad \implies \xi_t=-\frac{L}{2\lambda}(h_{z x}+\partial_{z} \xi_{x}+\partial_{x} \xi_{z}+\frac{\lambda}{z}\xi_x)\,.\nonumber
\end{align}
The last equation can be integrated by Fourier transforming all the functions along $t,x$ directions, thus reducing it to a first order differential equation in $z$ for the Fourier components.
\paragraph{Lightlike case.} In the null case, the differential equations in the components of $\xi_\mu$ read
\begin{align}
    0&= h_{++}+2\partial_+\xi_+\,,\nonumber\\
    0&=h_{z+}+\partial_+\xi_z+\partial_z\xi_++\frac2z\xi_+\,,\nonumber\\
    0&=h_{-+}+\partial_+\xi_-+\partial_-\xi_+-\frac1z\xi_z\,,
\end{align}
which can be integrated in the components of $\xi_\mu$ ,  by Fourier expanding all the functions along $x_\pm$. 
\paragraph{Timelike case.} Following the previous two paragraphs, the differential equations in the components of $\xi_\mu$ are found to be
\begin{align}
    0&= h_{tt}+2\partial_t\xi_t\,,\nonumber\\
    0&=h_{zt}+\partial_t\xi_z+\partial_z\xi_t + \frac{\lambda}{z}\xi_t +2\frac{\lambda}{L}\xi_x\,,\nonumber\\
    0&=h_{xt}+\partial_x\xi_t+\partial_t\xi_x-\frac2{ \lambda}{L}\xi_z\,,
\end{align}
which can be readily integrated by Fourier expanding all the functions along $t,x$.

\bibliographystyle{utphys}
\bibliography{refs.bib}

\end{document}